\documentclass[aip,pop,numerical,preprint]{revtex4-1}
\usepackage{CJK}
\usepackage{graphicx}
\usepackage{dcolumn}
\usepackage{bm}
\usepackage{amsmath}
\usepackage{amssymb}
\usepackage{color}

\begin{document}
\begin{CJK*}{GB}{gbsn}

\title{Numerical study of transition between even and odd toroidal Alfv\'en eigenmodes on EAST}


\author{Yawei Hou (ºîÑÅΡ)}
\affiliation{CAS Key Laboratory of Geospace Environment and Department of Engineering and Applied Physics, University of Science and Technology of China, Hefei, Anhui 230026, China}
\affiliation{KTX Laboratory and Department of Engineering and Applied Physics, University of Science and Technology of China, Hefei, Anhui 230026, China}

\author{Charlson C. Kim (½ðÖÇÉÆ)}
\affiliation{SLS2 Consulting, San Diego, California 92107, USA}

\author{Ping Zhu (Öìƽ)}\email{zhup@hust.edu.cn}
\affiliation{International Joint Research Laboratory of Magnetic Confinement Fusion
and Plasma Physics, State Key Laboratory of Advanced Electromagnetic
Engineering and Technology, School of Electrical and Electronic
Engineering, Huazhong University of Science and Technology, Wuhan, Hubei
430074, China}
\affiliation{Department of Engineering Physics, University of Wisconsin-Madison, Madison, Wisconsin 53706, USA}\affiliation{CAS Key Laboratory of Geospace Environment and Department of Engineering and Applied Physics, University of Science and Technology of China, Hefei, Anhui 230026, China}
\affiliation{KTX Laboratory and Department of Engineering and Applied Physics, University of Science and Technology of China, Hefei, Anhui 230026, China}

\author{Zhihui Zou (×ÞÖ¾»Û)}
\affiliation{CAS Key Laboratory of Geospace Environment and Department of Engineering and Applied Physics, University of Science and Technology of China, Hefei, Anhui 230026, China}
\affiliation{KTX Laboratory and Department of Engineering and Applied Physics, University of Science and Technology of China, Hefei, Anhui 230026, China}

\author{Youjun Hu (ºúÓÑ¿¡)}
\affiliation{Institute of Plasma Physics, Chinese Academy of Sciences, Hefei, Anhui 230031, China}

\author{Xingting Yan (ãÆÐÇÍ¢)}
\affiliation{CAS Key Laboratory of Geospace Environment and Department of Engineering and Applied Physics, University of Science and Technology of China, Hefei, Anhui 230026, China}
\affiliation{KTX Laboratory and Department of Engineering and Applied Physics, University of Science and Technology of China, Hefei, Anhui 230026, China}

\author{the NIMROD Team}


\date{\today}

\begin{abstract}
Linear simulations of toriodal Alfv\'en eigenmodes (TAEs) driven by energetic particles (EPs) on EAST (Experimental Advanced Superconducting Tokamak) are performed using the hybrid-kinetic MHD (HK-MHD) model implemented in NIMROD code. The EAST equilibrium is reconstructed using the EFIT code based on experimental measurement. The ¡°slowing down¡± distribution is adopted for modeling the equilibrium distribution of the energetic ions from the deuterium neutral beam injection on EAST. The frequency, the dominant poloidal mode number, the radial location and the detailed 2D mode structure of the TAE/RSAE/EPM modes are consistent between the eigenvalue analysis and the NIMROD simulation. As the $\beta$ fraction of EP increases, a transition between even and odd TAEs occurs, along with that between the ballooning and anti-ballooning mode structures. When the $\beta$ fraction of EP is close to the transition threshold, both types of TAEs coexist.
\end{abstract}


\maketitle
\end{CJK*}

\section{Introduction}
Since the velocity of energetic particles (EPs) is close to the phase velocity of Alfv\'en eigenmodes (AEs), EPs generated by heatings and fusion reactions in tokamak plasmas may excite toriodal Alfv\'en eigenmodes (TAEs), which can influence the stability and confinement of plasmas in burning regimes.\cite{Rosenbluth1975,Wong1999,Fasoli2007,Sharapov2013,Chen2016} It is necessary to study the physics of EP driven AEs in order to maintain the steady state of long-pulse plasma in presence of high power heating.

EAST(Experimental Advanced Superconducting Tokamak) is a medium-size tokamak with fully superconducting TF (Toroidal Field) and PF(Poloidal Field) coils, which has similar configuration to ITER (International Tokamak Experimental Reactor). The main design parameters are as follows: major radius $R = 1.7-1.9 m$, minor radius $a = 0.4-0.45 m$, toroidal magnetic field $B_t = 3.5 T$ and maximum plasma current $I_p = 1 MA$. After the upgrade of heating and current driving systems, especially the installation of the $4MW$ NBI system, the total auxilliary heating power of EAST has become more than $30MW$. For EAST NBI system, the power of each beamline is $2MW$ and the maximum injecting energy of Deuterium is 80 $keV$. In EAST discharge $\#48916$, the plasma energy for this eqilibrium is $127kJ$ and the energy stored in the energetic ions is $35kJ$. Hu et al \cite{Hu2014,Hu2016} has studied the linear feature of AEs for this discharge using the eigenvalue code GTAW\cite{Hu2014} and the kinetic-MHD code MEGA\cite{Hu2016}. To further examine the general features of AEs on EAST, we perform an eigen-analysis of AEs using the code AWEAC (Alfv\'en Wave Eigen-Analysis Code) and a linear simulation using the NIMROD code \cite{Sovinec2004, Kim2008} on the same EAST discharge.
It is found that the linear calculations results from NIMROD are consistent with the eigen-analysis results from AWEAC and GTAW.
The AE features of $n=2$ and $n=3$ modes from NIMROD are similar to those obtained from MEGA. Here $n$ is the toroidal mode number. However, for $n=4$ mode, a mode transition from even TAE to odd TAE has been revealed with the increase of $\beta$ fraction of EP, which is different from MEGA simulation.

According to the ideal MHD theory\cite{Cheng1986, Fu1995a,Berk1995,Fu1995b}, TAE is composed of two coupled poloidal harmonics, $m$ and $m+1$, for the same toroidal mode number $n$. If the two coupled poloidal harmonics have same sign, the formed TAE would be even TAE\cite{Fu1995b} which locates at the bottom end of TAE gap with ballooning mode structure. If the two coupled poloidal harmonics have opposite signs, the formed TAE would be odd TAE\cite{Berk1995} which locates at the top end of TAE gap with anti-ballooning mode structure. The existence of odd TAE, initially predicted from theory\cite{Berk1995}, was observed in JET experiment\cite{Kramer2004} with ICRH and LHCD heating. Both theory and experiment\cite{Fu1995a,Berk1995,Fu1995b,Kramer2004} suggest the even TAE is more robust than the odd TAE, which is also verified in our simulation.

The rest of the paper is organized as follows. Section II introduces the simulation model in NIMROD and section III introduces the eigenmode analysis method in AWEAC. In section IV, the NIMROD simulation setup is introduced. In section V, simulation results with different toroidal mode numbers, including mode structure, mode identification, EP $\beta$ fraction effect, are discussed. Finally, it comes to the summary and discussion in section IV.

\section{Simulation model in NIMROD}
The hybrid kinetic-MHD model implemented in the NIMROD code is used in our simulations. The background plasma and energetic ions are modeled using MHD equations and drift kinetic equations, respectively\cite{Kim2004}. The resistive two-fluid MHD equations are solved as an initial-boundary value problem that is decretized on a mesh of finite elements in the poloidal plane and with a finite Fourier series in the toroidal direction.\cite{Sovinec2004}
The hybrid kinetic-MHD model in NIMROD has been applied to the study of the fishbone mode in a model tokamak equilibrium, and the results have been benchmarked with the M3D-K code\cite{Fu2006} with good agreement\cite{Kim2008}. And this hybrid model has also been used to study the energetic particle effect on $n=1$ resistive MHD instability\cite{Takahashi2009, Brennan2012}, as well as the Alfv\'en Eigenmode\cite{Hou2018}.  For the sake of completeness of narrative and the convenience of reference, the main details of the model and its implementation are briefly outlined below.
The ideal MHD equations are as follows,
\begin{align}
\frac{\partial \rho}{\partial t} + \nabla\cdot(\rho\bm V) &= 0 ,\\
\rho\left(\frac{\partial \bm V}{\partial t}+\bm V\cdot\nabla\bm V\right) &=
\bm J\times\bm B-\nabla p_b-\nabla\cdot\bm P_f \label{eq:momen} ,\\
\frac{1}{\gamma-1}\left(\frac{\partial p}{\partial t} +
  \bm V\cdot\nabla p\right) &=
-p\nabla\cdot\bm V ,\\
\frac{\partial \bm B}{\partial t} &= -\nabla\times \bm E ,\\
\bm J &= \frac{1}{\mu_0}\nabla\times\bm B ,\\
\bm E + \bm V \times\bm B &= 0 ,
\end{align}
where subscripts $b,f$ denote bulk plasma and fast particles, $\rho,\bm V$ is fluid element density and velocity for the bulk plasma, neglecting the contribution of fast particles, $p$ the pressure of entire plasma, $p_b$ the pressure of bulk plasma, $\bm P_f$ the pressure tensor of fast particles, and $\gamma$ the ratio of specific heats, $\bm J$ the current density, $\bm B$ the magnetic field, $\bm E$ the electric field, and  $\mu_0$ the permeability of vacuum.

In HK-MHD model, it is assumed that the number density of fast species $n_f$ is much lower than that of bulk plasmas $n_b$ but the fast species pressure $p_f$ is on the same order of the bulk plasma pressure $p_b$, i.e. $n_f \ll n_b$
and $\beta_f\sim \beta_b$, and $\beta\equiv 2\mu_0p/B^2$
is the ratio of thermal energy to magnetic energy. In this approximation,
we neglect the contribution of energetic particles to the center of mass velocity.
If we take the center of the mass velocity of energetic ions to be zero,  $\bm P_f$ in the momentum equation can be calculated from the distribution function and the velocity of energetic ions,
\begin{equation}
  \bm P_f = m_f \int \bm v_f \bm v_f f_f(\bm r_f,\bm v_f) d^3 v_f \label{eq:fmom},
\end{equation}
where $m_f$, $\bm r_f$ and $\bm v_f$ are the mass, the spatial coordinate vector and the velocity of fast ions, respectively.\\

The $\delta f$ PIC method is used to solve the drift kinetic equation of energetic particles.
In the limit of strong magnetic field, the drift kinetic approximation
reduces the 6D phase space $(\bm r,\bm v)$ to 5D $(\bm r, v_{\parallel},\mu)$
with one adiabatic invariant (i.e. the first adiabatic invariant
$\mu = \frac 12 mv_{\bot}^2/B$).
If we substitute $f_f=f_{f0}+\delta f_f$
into Eq. \eqref{eq:fmom}, where $f_{f0}$ and $\delta f_f$ are the equilibrium
and the perturbed distribution function of fast particles, respectively, then $\bm P_f$ can be calculated as following
\begin{equation}
  \bm P_f =\bm P_{f0} + \delta\bm P_f ,
\end{equation}
\begin{equation}
  \delta\bm P_f=m_f \int \bm v_f \bm v_f \delta f_f(\bm r_f,\bm v_f) d^3 v_f ,
\end{equation}
where $\bm P_{f0}$ and $\delta\bm P_f$ are the equilibrium and the perturbed fast particle pressure tensor, respectively.
The condition for the force balance in equilibrium is given by
\begin{equation}
  \bm J_0\times\bm B_0 = \nabla p_0 + \nabla p_{f0} ,
\end{equation}
where the assumption is that the anisotropic components of fast particle
pressure tensor in equilibrium are zero and the tensor $\bm P_{f0}$ is reduced to a scalar
$p_{f0}$. Note that the steady state fields satisfy a scalar pressure force
balance, which is based on the assumption that the form of equilibrium energetic particle distribution
is isotropic in velocity space. With the solution for $\delta f_f$, we can calculate the pressure tensor. In the drift-kinetic approximation, the CGL-like pressure tensor can be used, $\delta \bm P_f = \delta p_{\bot} \bm I + (\delta p_{\parallel}-\delta p_{\bot}) \bm b\bm b$, where $\delta p_{\bot} = \int \mu B\delta f_f d^3 v_f$, $\delta p_{\parallel} = \int v_{\parallel}^2\delta f_f d^3v_f$, $\bm I$ is the unit tensor, and $\bm b=\bm B/B$.

The slowing down distribution function is used for the
energetic ions,
\begin{equation}
  f_0 = \frac{P_0 \exp((P_{\zeta}/\psi_n))}{\varepsilon^{3/2}+\varepsilon_c^{3/2}} ,
\end{equation}
where $P_0$ is a normalization constant, $\varepsilon$ the particle energy,
$\varepsilon_c$ the critical slowing down energy, $P_{\zeta}=g\rho_{\parallel}-\psi_p$ is the
canonical toroidal momentum, $g=RB_{\phi}$, $\rho_{\parallel}=mv_{\parallel}/ qB$, $\psi_p$ is
the poloidal flux, and $\psi_n=C\psi_0$, where $\psi_0$ is the total flux and $C$ is a constant parameter used to match the equilibrium pressure profile.
This distribution function models the slowing down of a monoenergetic beam of ions or fusion alpha particles where the
collisions are predominantly with the background electrons.

\section{The eigenmode analysis method in AWEAC}
AWEAC is developed to solve the ideal MHD eigenmode equations \cite{Cheng1986,Hu2014} with python and provide the radial mode structure and the spectrum of AEs. AWEAC can input equilibrium generated using EFIT (Fig.\ \ref{Fig1} (a)), and transform the cylindrical coordinates $(R,\phi, Z)$ to flux coordinates $(\psi,\theta,\zeta)$ (Fig.\ \ref{Fig1} (b)), where $\psi=(\Psi-\Psi_{axis})/(\Psi_{LCFS}-\Psi_{axis})$ is the normalized poloidal flux with $\Psi$ being poloidal flux, $\Psi_{axis}$ and $\Psi_{LCFS}$ the poloidal fluxes at magnetic axis and last closed flux surface (LCFS), $\theta$ the equal-arc length poloidal angle, and toroidal angle $\zeta$ is picked to make the magnetic field line straight in the plane $(\theta,\zeta)$ \cite{Hu2014}. After reading in the equilibrium from EFIT, AWEAC first finds out the LCFS, which is the red curve shown in Fig.\ \ref{Fig1} (a), for example. In the domain within LCFS, AWEAC can generate grid points in flux surface coordinates based on the grid points in cylindrical coordinates from the EFIT eqilibrium. The blue lines and red lines in Fig.\ \ref{Fig1} (b) schematically represent the flux surfaces and the equal-arc-length poloidal angles from a finer grid used in AWEAC. \\

The continuous spectrum equations\cite{Cheng1986} from the linearized ideal MHD model are
\begin{equation}
\left[
\begin{matrix}
E_{11}&E_{12} \\
E_{21}&E_{22}
\end{matrix}
\right]
\left[
\begin{matrix}
\xi_{s}\\
\nabla \cdot \bm \xi
\end{matrix}
\right]
=0,
\end{equation}
where $\bm \xi$ and $\xi_{s}$ are the plasma displacement vector and its poloidal component, respectively.

The matrix elements are
\begin{equation}
E_{11} = -\frac{\omega^2\rho_0\lvert\nabla\Psi\rvert^2}{B_{0}^2} -\mu_0^{-1}\bm B_0\cdot\nabla(\frac{\lvert\nabla\Psi\rvert^2}{B_{0}^2}\bm B_0\cdot\nabla)
\end{equation}
\begin{equation}
E_{12}=-2\kappa_s\gamma p_0
\end{equation}
\begin{equation}
E_{21}=2\mu_0^{-1}\kappa_s
\end{equation}
\begin{equation}
E_{22}=\frac{\mu_0^{-1}B_0^2+\gamma p_0}{B_0^2}+\frac{\gamma p_0}{\mu_0\omega^2\rho_0}\bm B_0\cdot\nabla(\frac{\bm B_0\cdot\nabla}{B_{0}^2}),
\end{equation}
where the subscript "0" denotes the equilibrium quantities, $\bm B_0$ and $B_0$ are the vector and magnitude of magnetic field, $\rho_0$ the mass density, $p_0$ the plasma thermal pressure, $\gamma$ the ratio of specific heats and picked to be $5/3$ in our calculation, $\mu_0$ the vacuum permeability, $\kappa_s$ the geodesic curvature, $\omega$ the frequency of perturbation.

By solving equation at each magnetic surface, the eigenvalue $\omega$ as function of $\psi$ can be found from $Det|E(\omega)|=0$.
In particular, by multiplying $\omega^2$ to matrix elements $E_{21}$ and $E_{22}$, the matrix can be written as
\begin{equation}
E=E_a+\omega^2E_b
\end{equation}
and the continuous spectrum equation can be written as
\begin{equation}
E_a
\left[
\begin{matrix}
\xi_{s}\\
\nabla \cdot \bm \xi
\end{matrix}
\right]
=-\omega^2E_b
\left[
\begin{matrix}
\xi_{s}\\
\nabla \cdot \bm \xi
\end{matrix}
\right],
\end{equation}
where
the elements of matrix $E_a$ and $E_b$ are
\begin{equation}
E_{11}^a = -\mu_0^{-1}\bm B_0\cdot\nabla(\frac{\lvert\nabla\Psi\rvert^2}{B_{0}^2} \bm B_0\cdot\nabla)
\end{equation}
\begin{equation}
E_{12}^a=-2\kappa_s\gamma p_0
\end{equation}
\begin{equation}
E_{21}^a=0
\end{equation}
\begin{equation}
E_{22}^a=\frac{\gamma p_0}{\mu_0\rho_0}\bm B_0\cdot\nabla(\frac{\bm B_0\cdot\nabla}{B_{0}^2})
\end{equation}
\begin{equation}
E_{11}^b = -\frac{\rho_0\lvert\nabla\Psi\rvert^2}{B_{0}^2}
\end{equation}
\begin{equation}
E_{12}^b=0
\end{equation}
\begin{equation}
E_{21}^b=2\mu_0^{-1}\kappa_s
\end{equation}
\begin{equation}
E_{22}^b=\frac{\mu_0^{-1}B_0^2+\gamma p_0}{B_0^2}.
\end{equation}

Since $E_{22}^a/E_{22}^b$  $\approx$ $(\gamma \beta/2)/[1+\gamma \beta/2]$, when $\beta \ll 1$, the term $E_{22}^a$ can be dropped, which is called the slow sound approximation\cite{Chu1992,Deng2012}. This approximation will remove the sound continua while keeping the Alfv\'en continua nearly unchanged.
In the following calculation, we adopt the slow sound approximation to render the Alfv\'en continua more clear.

Fourier transform is further applied for an arbitrary perturbation $A$ over $\theta$ and $\zeta$ direction
\begin{equation}
A(\psi,\theta,\zeta,t)= \sum_{n=-\infty}^{\infty} \sum_{m=-\infty}^{\infty} A_{nm}(\psi,t)e^{i(m\theta-n\phi)},
\end{equation}
 where the $A_{nm}$ is given by
\begin{equation}
A_{nm}(\psi,t)= \frac{1}{(2\pi)^2} \int_{0}^{2\pi} \int_{0}^{2\pi} A(\psi,\theta,\zeta,t)e^{i(n\phi-m\theta)},
\end{equation}
where $t$ is time, $n$ is the toroidal mode number, and $m$ the poloidal mode number.
The spectrum equation is then solved in the Fourier space of (m,n).

\section{Simulation setup}
\begin{figure}
\includegraphics[height=9cm,width=6.5cm]{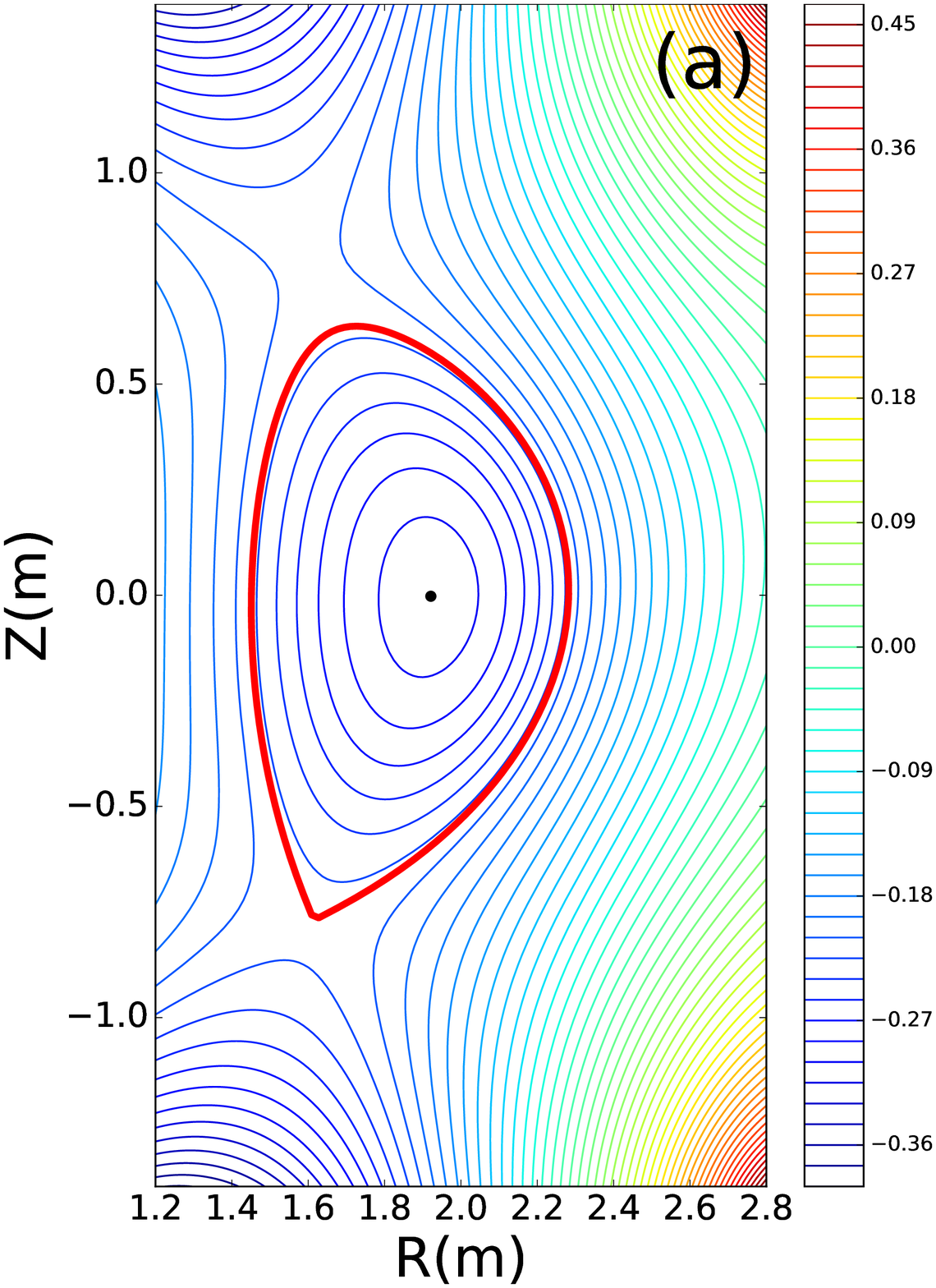}
\includegraphics[height=9cm,width=5.3cm]{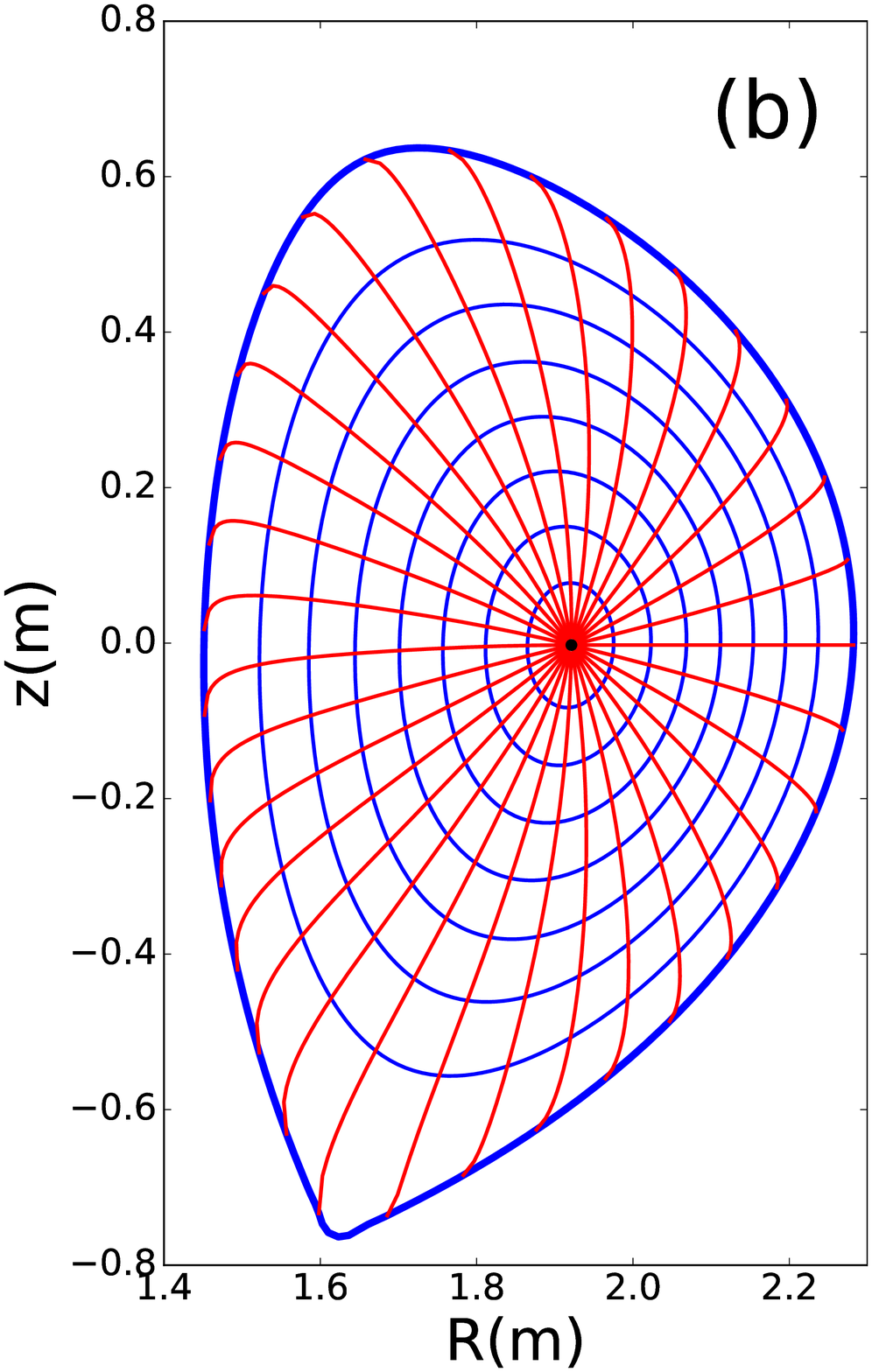}
\caption{(a) Contour plot of equilibrium poloidal flux of background plasma in $(R,Z)$ coordinate. The red curve represents the last closed flux surface (LCFS). (b) Diagram of the mesh grids of flux coordinate used in the Alfv\'en continuum calculation.
Poloidal equal-arc length are used in this flux coordinate. The black point locates at magnetic axis.
The blue lines are flux surfaces with same poloidal flux.
The red lines from the magnetic axis represents the grids with same poloidal angle. The mesh grids used in the real calculation are much finer than the diagram.  \label{Fig1}}
\end{figure}

The equilibrium is reconstructed using the EFIT code with experimental data from EAST discharge \#$48916$ at $4.5 s$. The flux surfaces of the equilibrium in $(R,Z)$ coordinate and the mesh grid based on the magnetic flux coordinate are shown in Fig.\ \ref{Fig1}. A double-null configuration can be identified in this equilibrium and the LCFS is connected to the lower X point. As can be seen from the LCFS, the equilibrium is up-down asymmetric. There is $127kJ$ energy stored in plasma for this equilibrium.

\begin{figure}
\includegraphics[height=5cm,width=6cm]{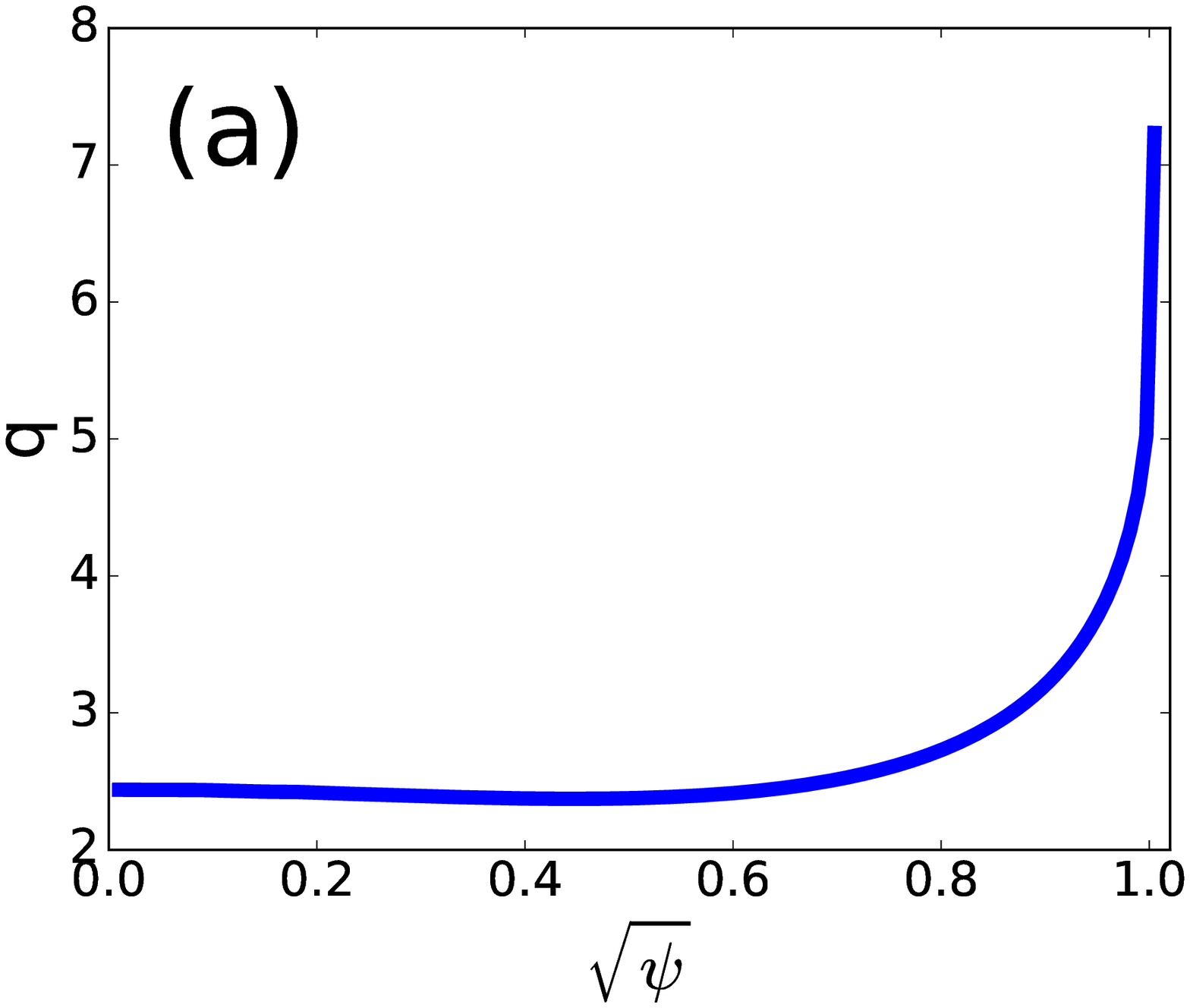}
\includegraphics[height=5cm,width=6cm]{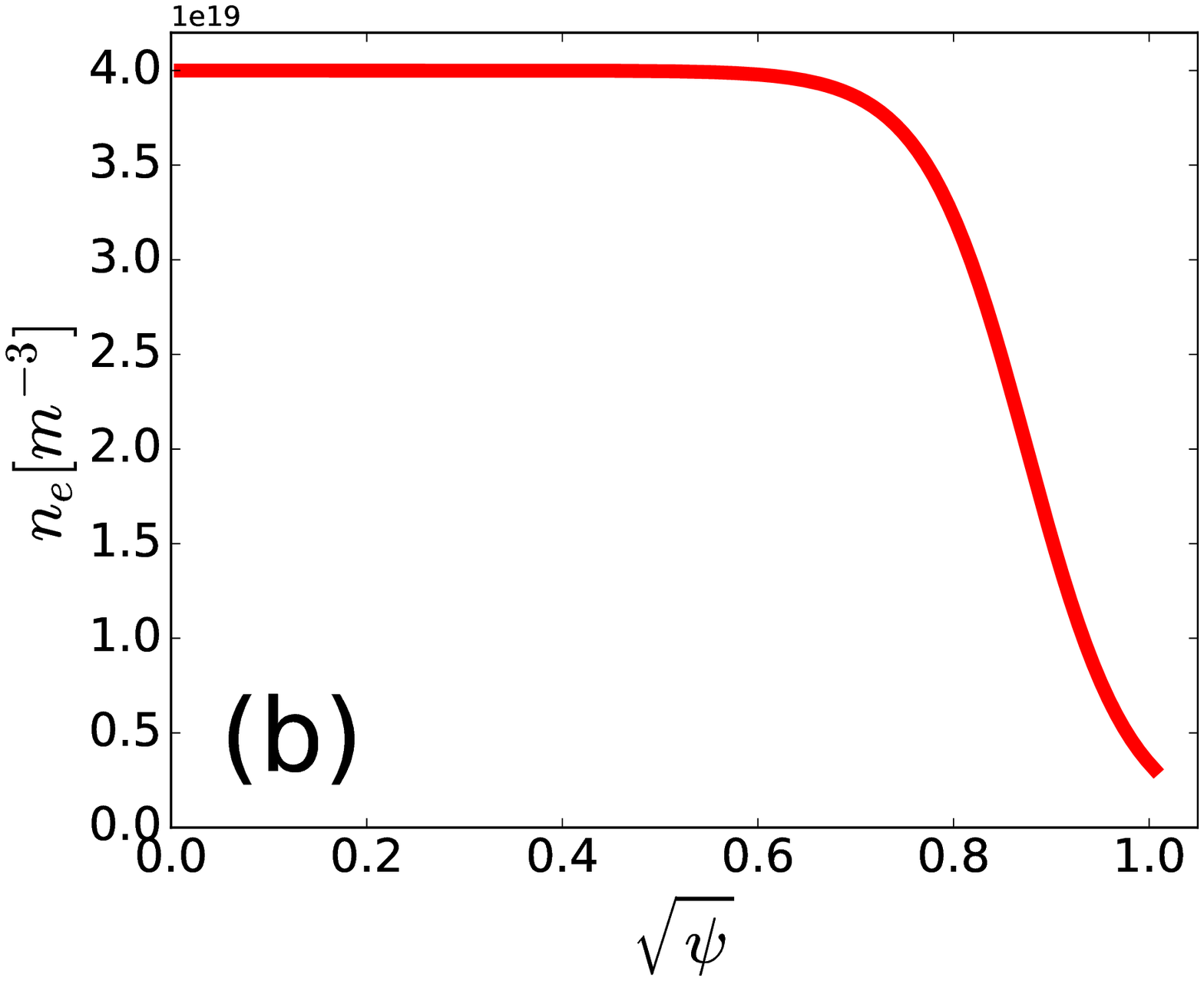}
\caption{Radial profile of safety factor (a) and electron density (b) of background plasma.
\label{Fig2}}
\end{figure}

The safety factor $q$ profile and electron density profile of background plasma are shown in Fig.\ \ref{Fig2}. Deuterium plasma is used in the discharge, where the mass density $\rho=n_{e}m_D$ for the plasma. At the magnetic axis, the safety factor is $q_0=2.438$. At the radial position $\sqrt{\psi}=0.4$, safety factor reaches to the minimum value $q_{min}=2.273$. In the region $\sqrt{\psi}\leq 0.4$, there is a weak reverse shear for the safety factor. At the magnetic flux surface of $95\%$ ploidal magnetic flux, the safety factor is $q_{95}$.

 The profile of the energetic particles loaded in the physical space is proportional to $\exp(-\psi_p/\psi_n)$. In order to maintain the force balance, parameters $P_0$ and $C$ in the slowing down distribution function can be calculated from the fitting of equilibrium pressure profiles of the background plasma. The fitting value $\psi_n$ used in NIMROD simulation is $0.088$, which is smaller than the value $0.4$ used in MEGA simulation\cite{Hu2016}. This can make the EP pressure gradient greater in NIMROD simulation than MEGA simulation\cite{Hu2016}.

 In the velocity space, energetic particles are loaded isotropically according to $1/(\varepsilon^{3/2}+\varepsilon_c^{3/2})$, where $\varepsilon$ and $\varepsilon_c$ are the energy and the critical energy of energetic particle respectively. There exists a critical velocity of energetic ions $v_c$, at which the collisional frictions of energetic ions with thermal electrons and ions are equal. In this case, with the NBI of deuterium, the critical energy can be written as $\varepsilon_c=\frac{1}{2} m_D v_c^2$. From the interaction model of beam and thermal plasma, the critical velocity can be calculated by
 \begin{equation} \label{eq:v_c}
  v_c = (\frac{m_e}{m_i}\frac{3\sqrt{\pi}}{4})^{1/3} v_{the},
\end{equation}
where the $m_e$ and $m_i$ are the mass of thermal electrons and thermal ions respectively, and $v_{the}$ is the thermal velocity of electrons.
In this shot, the characteristic electron temperature is considered to be $T_e=2 keV$, so the corresponding critical velocity should be $v_c=1.89 \times 10^6 m/s$ from the equation\ \eqref{eq:v_c}.
In order to model the NBI, we set the beam velocity $v_b=2.35 \times 10^6 m/s$, which corresponds to $58keV$ of the kinetic energy of deuterons. The setting of slowing dowm model is consistent with MEGA simulation\cite{Hu2016}, even though the cutoff width and the pitch angle effect of beams are not included in NIMROD simulation.

\begin{figure}
\includegraphics[height=5cm,width=6cm]{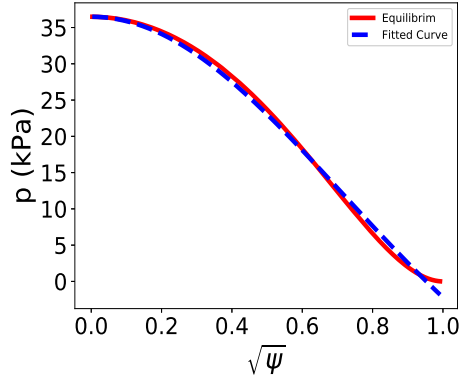}
\caption{Fitting of the pressure profile of background plasma to get the parameters used in the slowing distribution of energetic particles.
The red solid line is the pressure profile from equilibrium. The blue dotted line is the fitted curve. \label{Fig3}}
\end{figure}

In NIMROD simulation, since the modes we study are all core modes inside the pedestal, it is reasonable to use fixed boundary without vacuum region. The initial perturbation of magnetic field is set to be $\delta B/B_0$, which $B_0$ is the equilibrium magnetic field. The time step of the MHD evolution is set to be $10^{-7}$ s, and the time step of energetic particle evolution base on $\delta f$ PIC method is set to be $1/30-1/50$ of MHD time step.

\section{SIMULATION RESULTS}
We first benchmark the NIMROD results with those from both eigen-value analysis code (AWEAC and GTAW \cite{Hu2016})  and hybrid kinetic-MHD code (MEGA\cite{Hu2016}). It is found that the NIMROD results for the toroidal mode numbers $n=2,3,5,6$ are consistent with those from AWEAC, GTAW and MEGA. But for the $n=4$ case, the NIMROD results show mode transition from even TAE to odd TAE due to enhanced driving of energetic particles.

\subsection{AEs/EPMs for $n=2,3,5,6$}
\subsubsection{Mode structure}
\begin{figure}
\includegraphics[height=5cm,width=4cm]{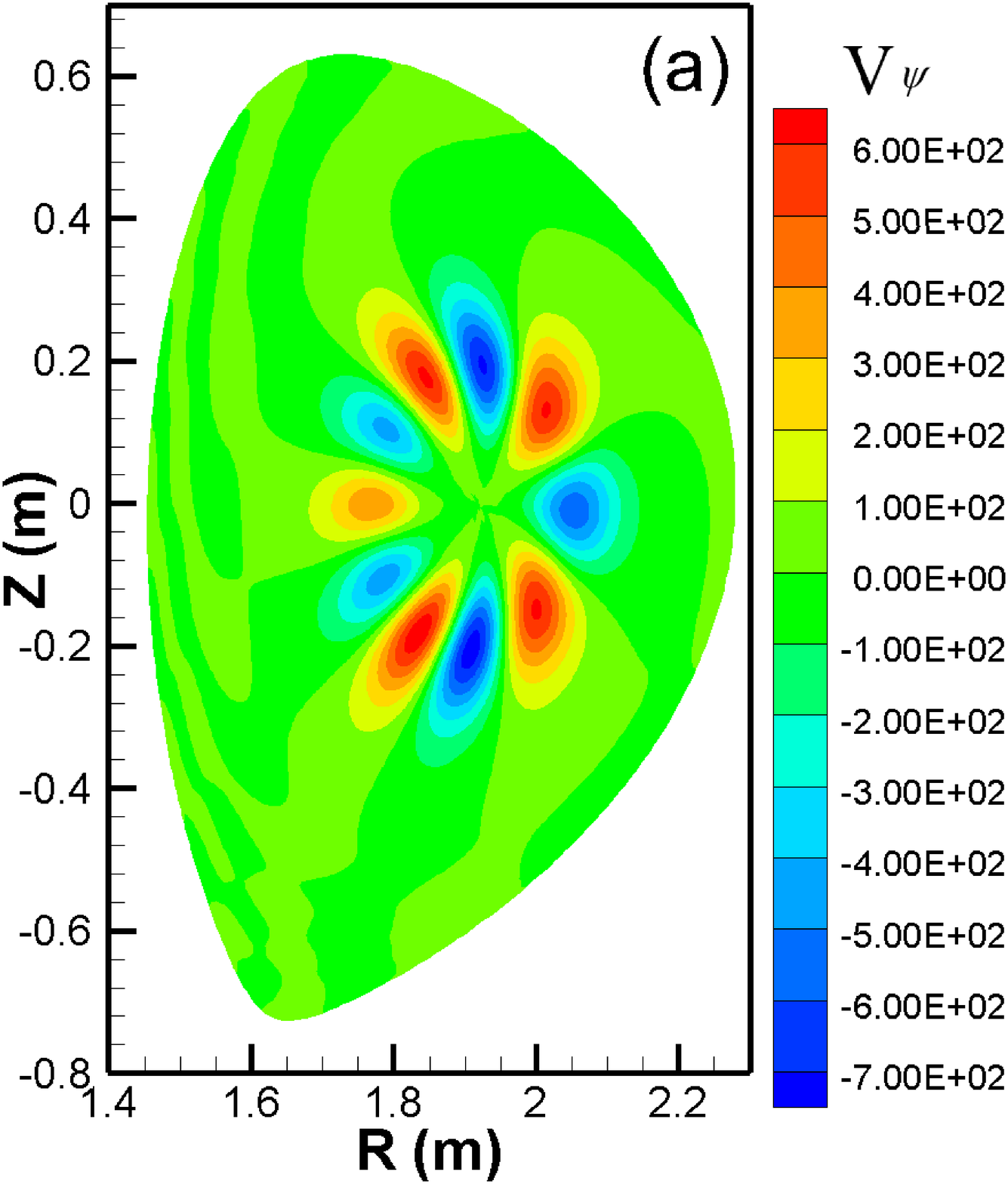}\includegraphics[height=5cm,width=4cm]{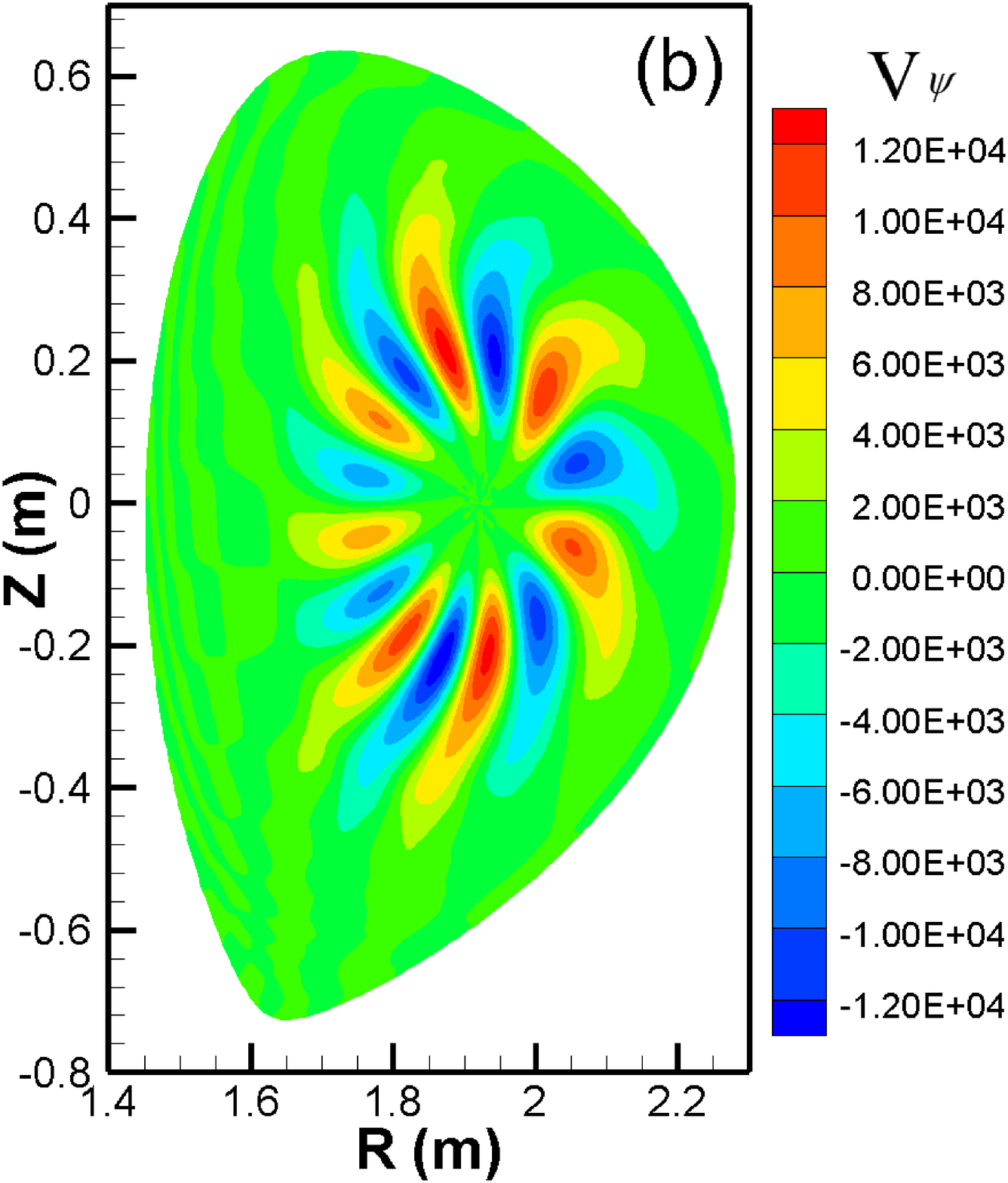}
\includegraphics[height=5cm,width=4cm]{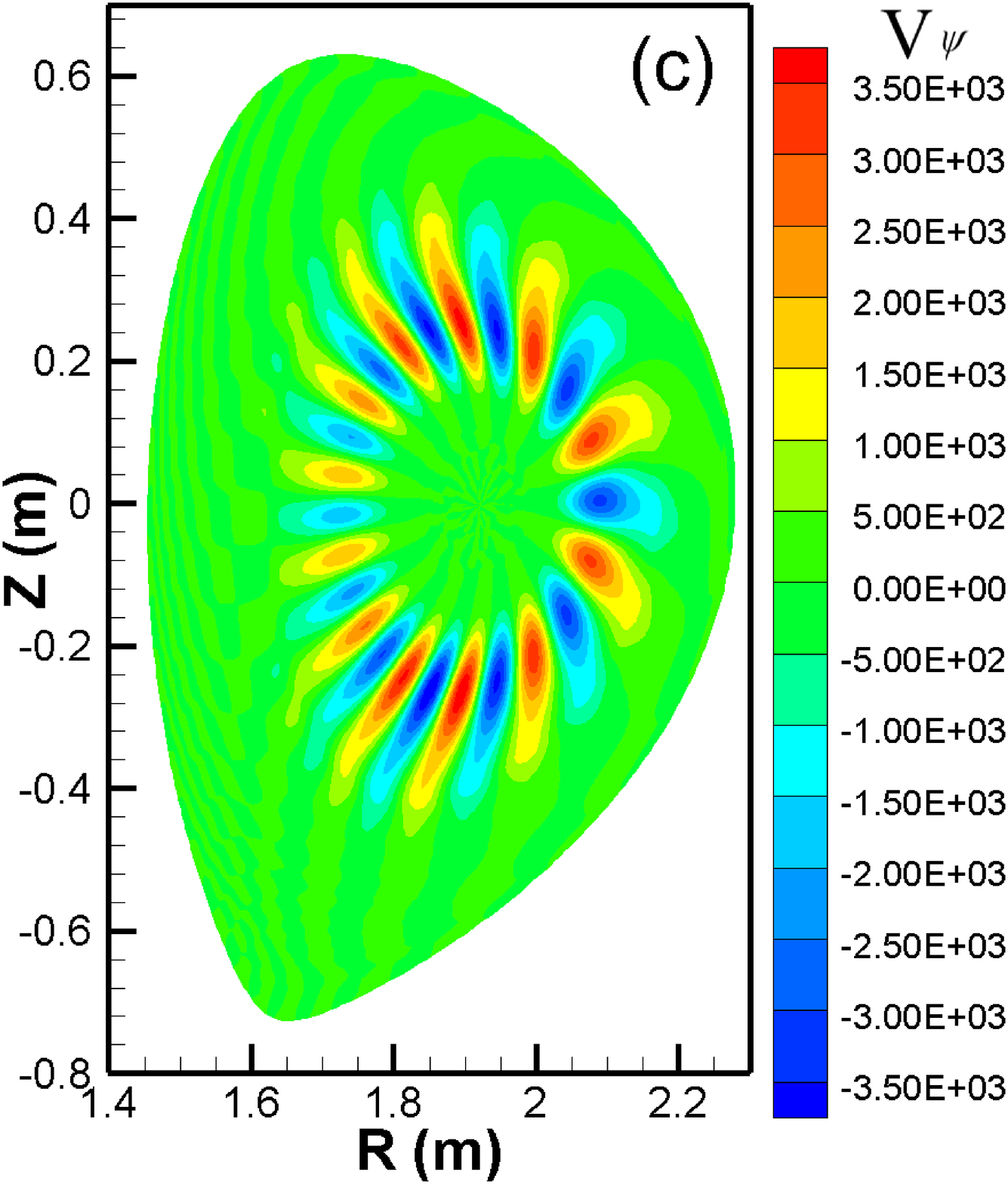}\includegraphics[height=5cm,width=4cm]{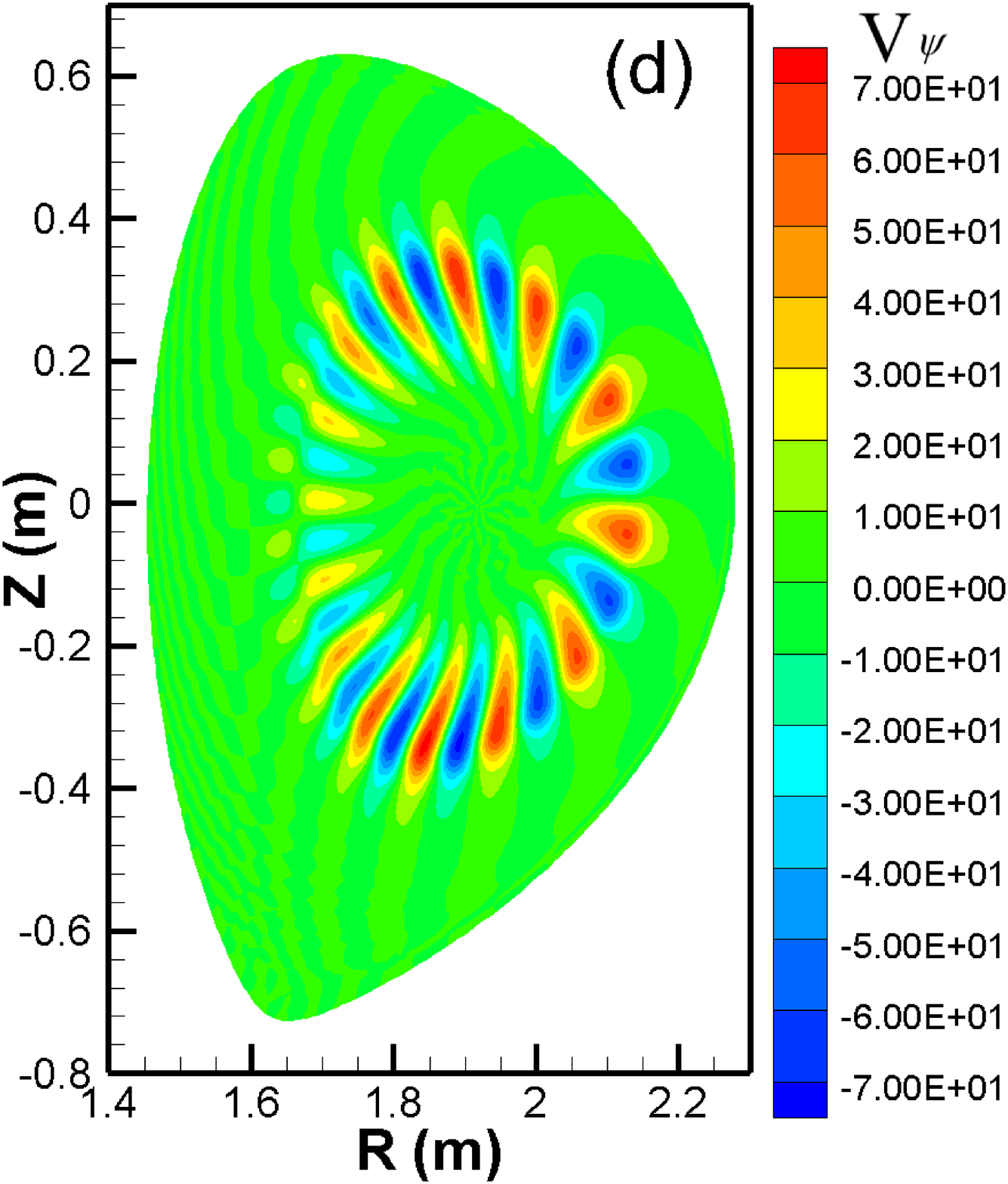}
\caption{Contour plot of the radial component of velocity of background plasma with $\beta_f=0.2835$. (a) n=2, (b) n=3, (c) n=5, (d) n=6. \label{Fig4}}
\end{figure}

$2D$ contour plots of the radial velocity of plasma are shown in Fig.\ \ref{Fig4}. The $n=2$ mode is located in the region $0.2 < \sqrt{\psi} < 0.5$. The center of the modes of $n=3,5,6$ are located around $\sqrt{\psi}=0.5$ region. The main poloidal harmonics of the modes for $n=2,3,5,6$ are  $m=5, 7, 12, 14$, respectively.

\subsubsection{Mode identification}
\begin{figure}
\includegraphics[height=6cm,width=8cm]{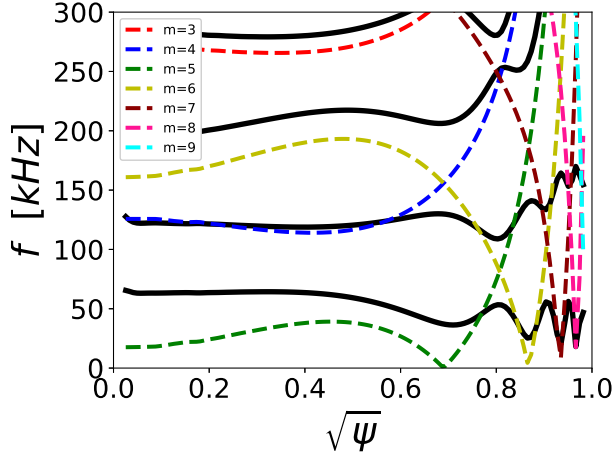}
\caption{(a) Alfv\'en continua with toroidal mode number $n=2$ calculated based on the slow-sound approximation from AWEAC (solid lines). Alfv\'en continua from the cylindrical geometry limit (dotted lines) are also given. \label{Fig5}}
\end{figure}

In order to identify the mode nature for $n=2$ case, we take Alfv\'en continua into consideration. From the NIMROD simulation, we get the mode frequency $70.8kHz$, which is consistent with the value $69kHz$ from AWEAC and GTAW. The corresponding growth rate is $2.395\times 10^4 s^{-1}$ from NIMROD simulation. Because the radial location of $n=2$ mode is $0.2 < \sqrt{\psi} < 0.5$, this mode is located in the TAE gap shown in the Fig.\ \ref{Fig5}. As can be seen in Fig.\ \ref{Fig5}, Alfv\'en continua of $m=4$ and $m=5$ in the cylindrical limit are well separated from each other, so there is only a weak coupling between these two poloidal harmonics in the toroidal geometry. This mode should not be the TAE mode. Considering that the safety factor reaches the minimum at $\sqrt{\psi}=0.4$, which is also the location region of this mode, the mode should be RSAE (Reverse Shear Alfv\'en Eigenmode).

\begin{figure}
\includegraphics[height=6cm,width=8cm]{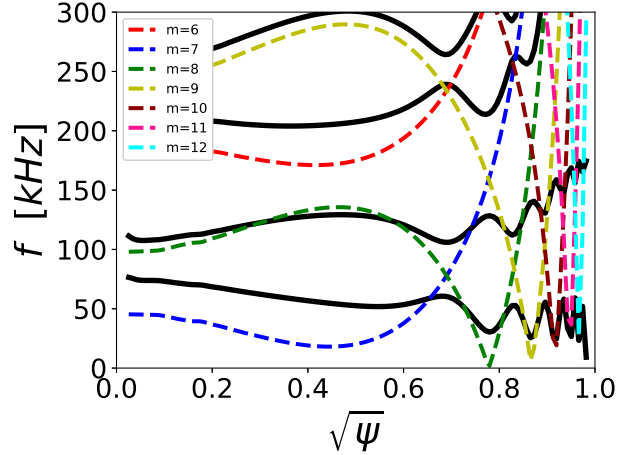}
\caption{Alfv\'en continua with toroidal mode number $n=3$ calculated based on the slow-sound approximation from AWEAC (solid lines). Alfv\'en continua from the cylindrical geometry limit (dotted lines) are also given. \label{Fig6}}
\end{figure}

For $n=3$ case, Alfv\'en continua are plotted in the Fig.\ \ref{Fig6}, together with the continua of $m=7$ and $m=8$ poloidal harmonics in the cylindrical limit. The mode frequency from NIMROD is $68.56kHz$, which is consistent with frequency ($69kHz$) from AWEAC, GTAW and MEGA. The growth rate is $2.6023\times 10^4 s^{-1}$, which is larger than that in $n=2$ case with the same EP $\beta$ fraction. From Fig.\ \ref{Fig4} (b), one can see the mode is located in the radial region $0.2 < \sqrt{\psi} < 0.6$. The mode touches the bottom line of TAE gap, which is induced by the coupling of $m=7$ and $m=8$ poloidal harmonics (Fig.\ \ref{Fig6}). Because $m=7$ harmonic is dominant in the $2D$ structure and the mode intersects more strongly with $m=7$ than with $m=8$ harmonics, this mode can be identified to be EPM.

\subsubsection{EP $\beta$ fraction effect}
\begin{figure}
\includegraphics[height=3.5cm,width=4cm]{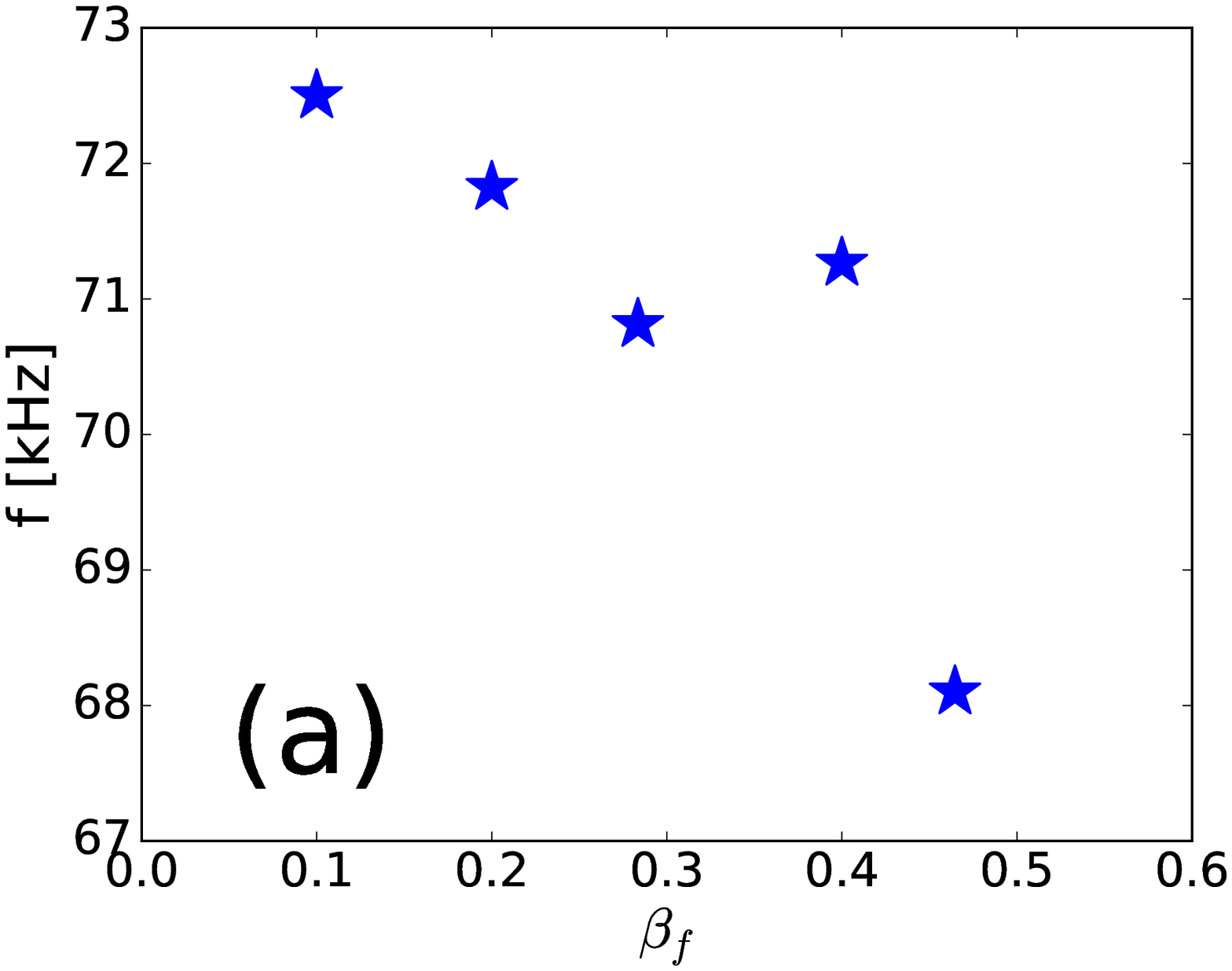}
\includegraphics[height=3.5cm,width=4cm]{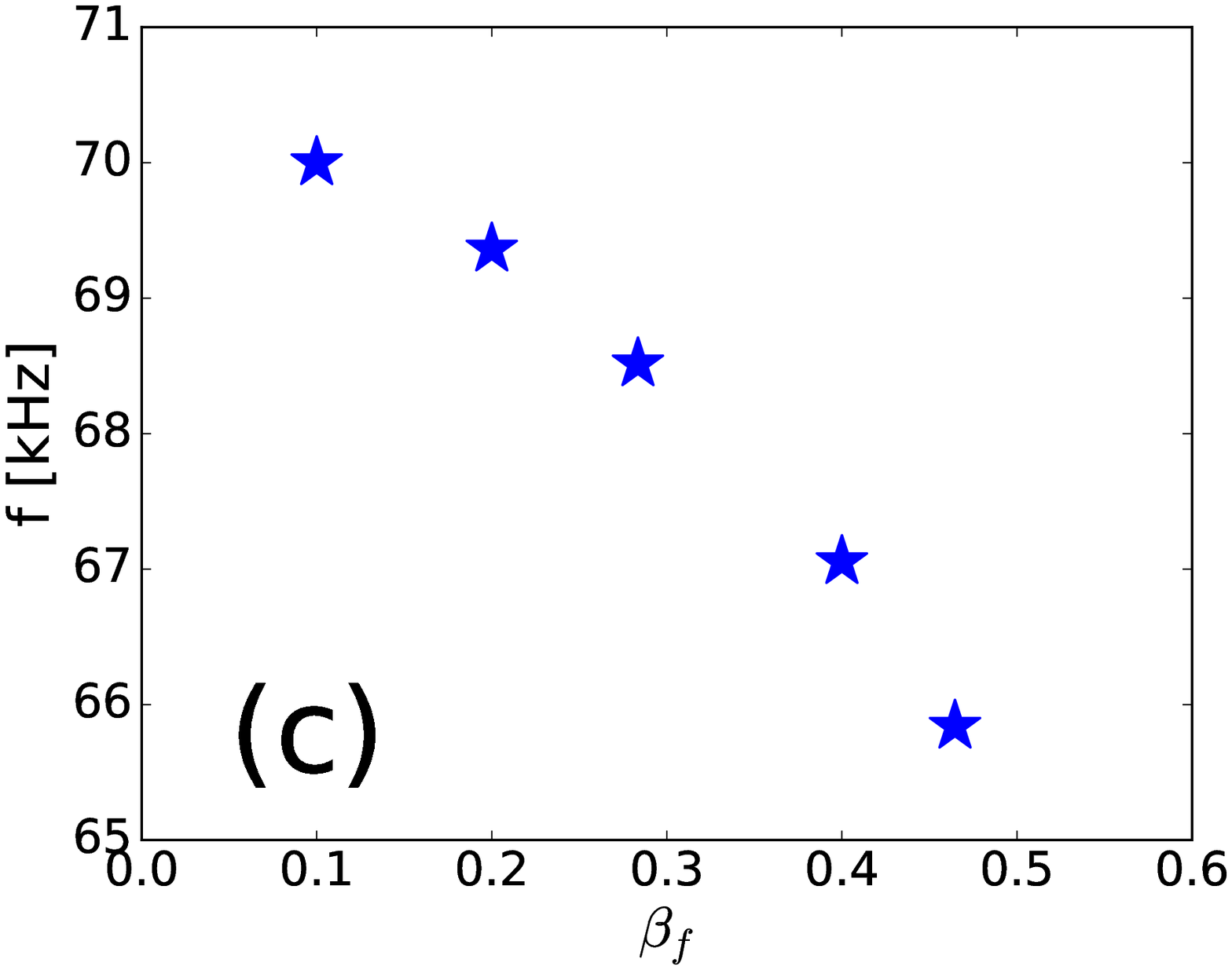}
\includegraphics[height=3.5cm,width=4cm]{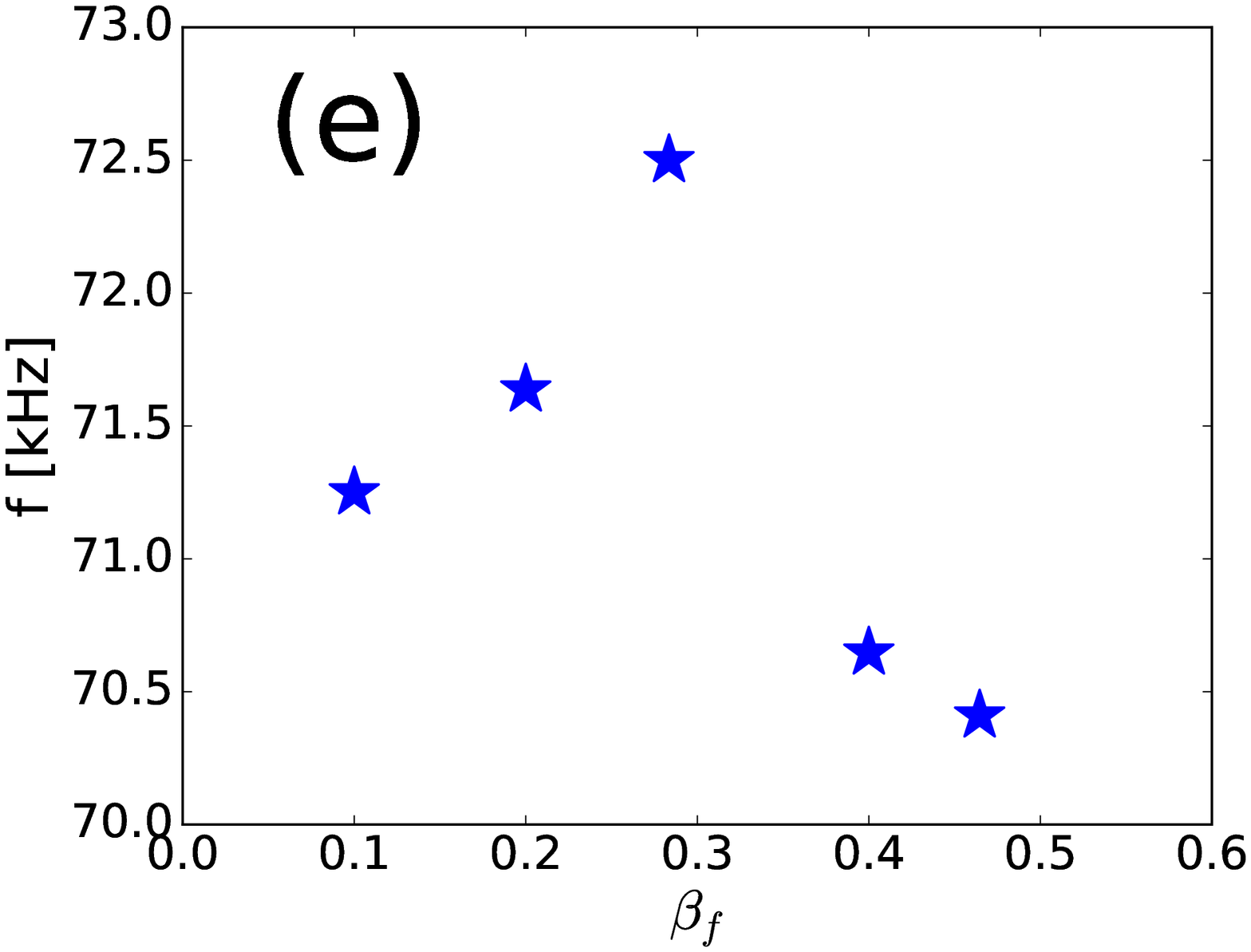}
\includegraphics[height=3.5cm,width=4cm]{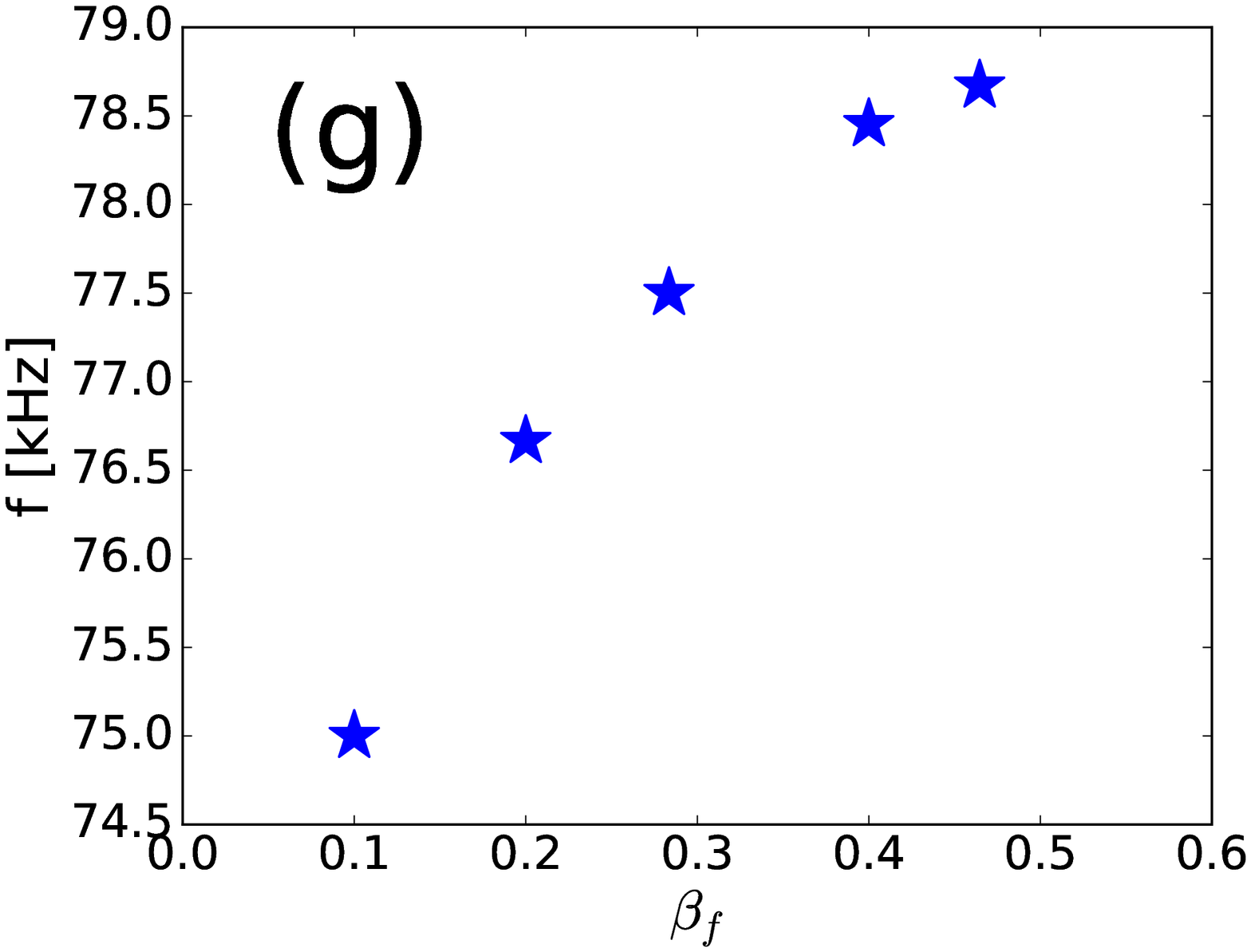}
\includegraphics[height=3.5cm,width=4cm]{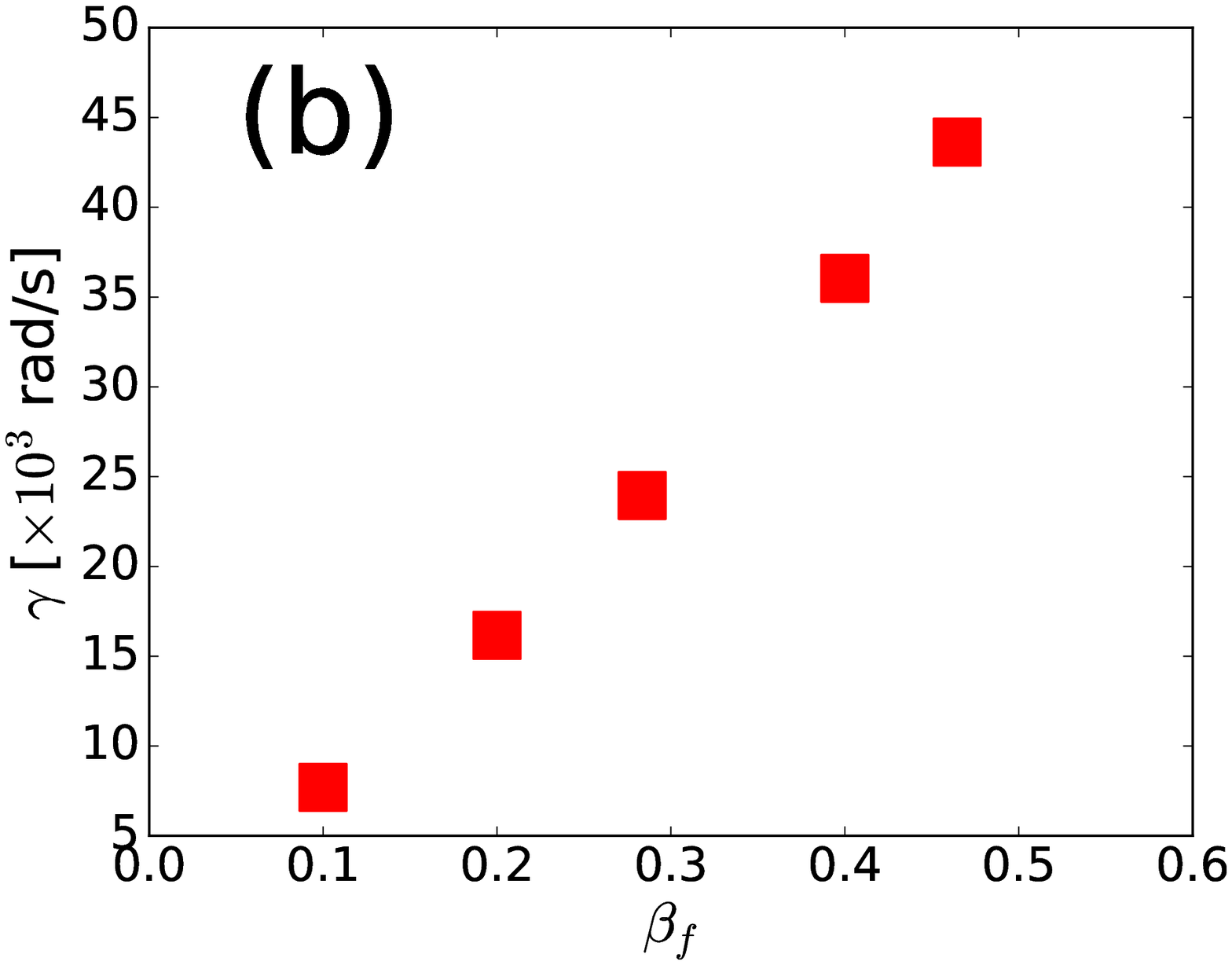}
\includegraphics[height=3.5cm,width=4cm]{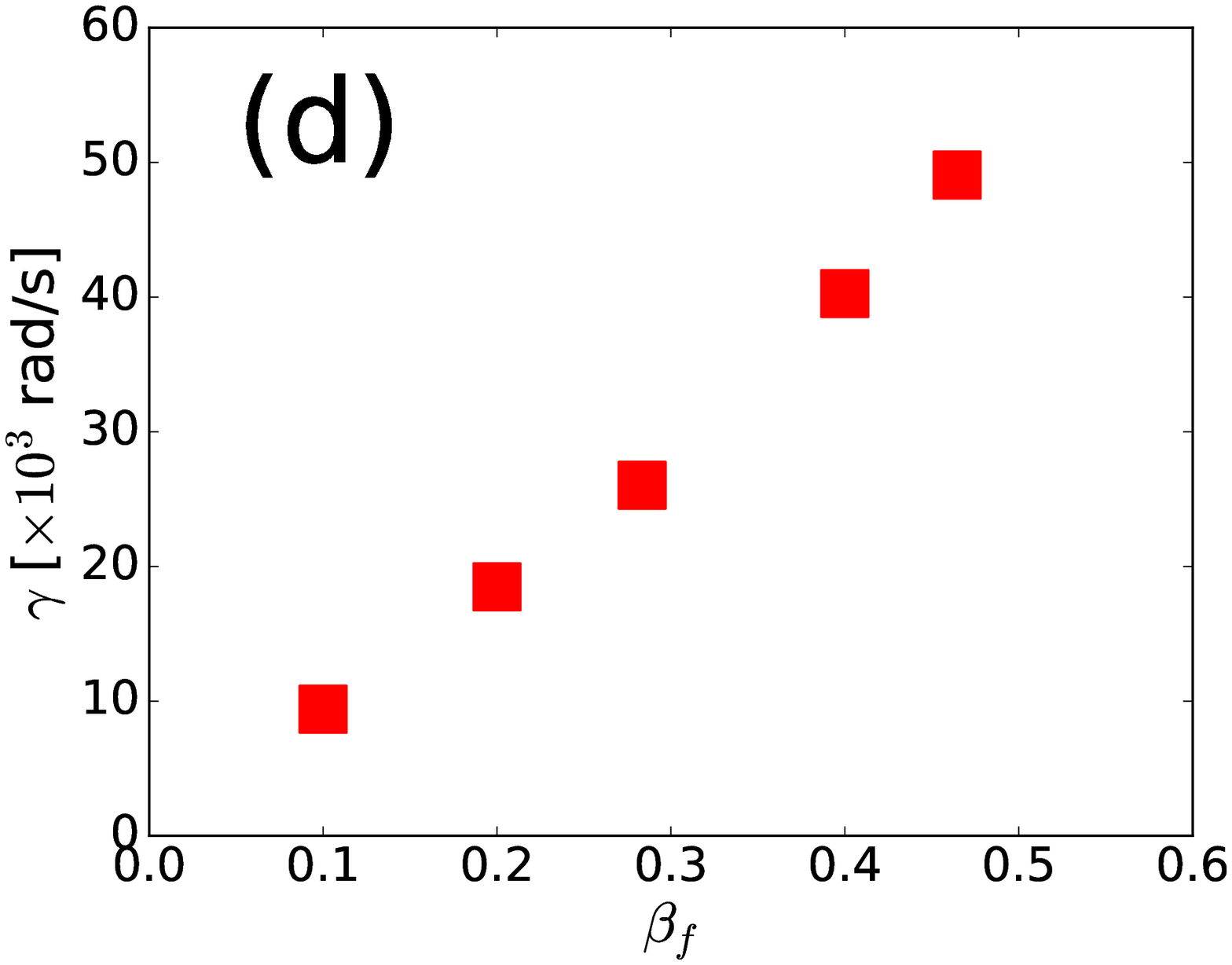}
\includegraphics[height=3.5cm,width=4cm]{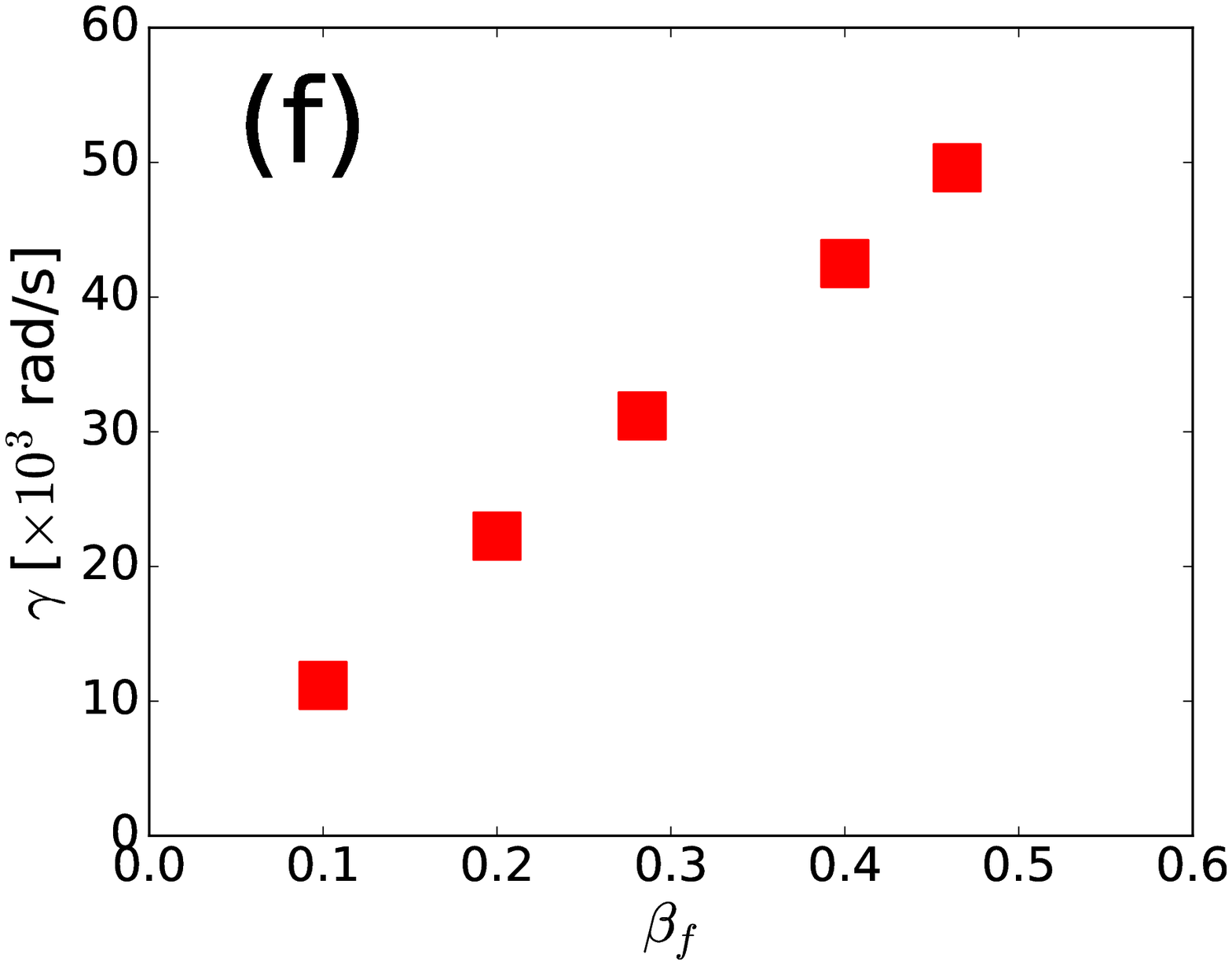}
\includegraphics[height=3.5cm,width=4cm]{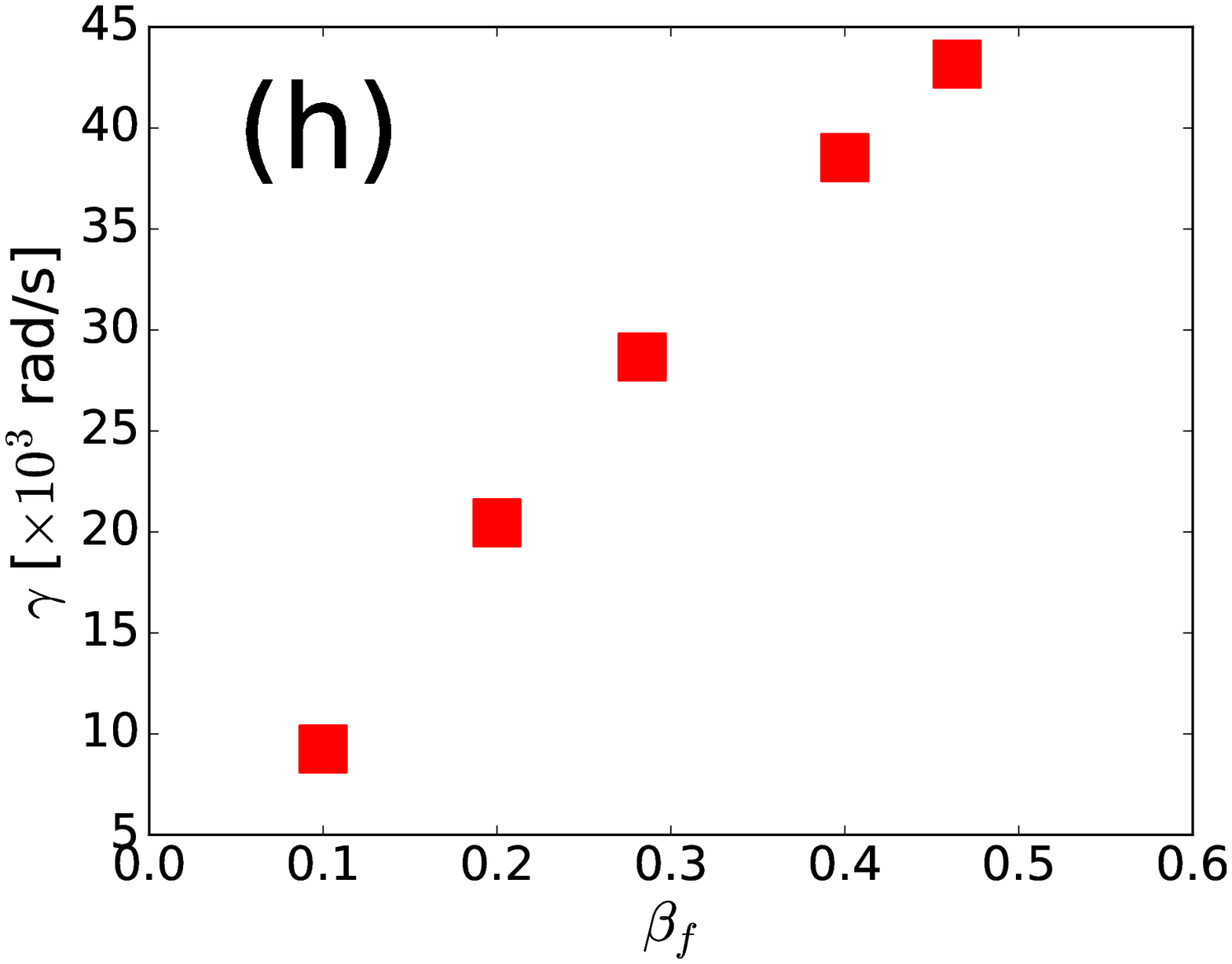}
\caption{The frequency (blue star)
and growth rate (red square) of TAE mode as functions of the EP $\beta_f$ for $n=2 \ \ ((a) \ \ and \ \ (b)), \ \ 3 \ \ ((c) \ \ and \ \ (d)), \ \ 5 \ \ ((e) \ \ and \ \ (f)), \ \ 6 \ \ ((g) \ \ and \ \ (h))$ cases. \label{Fig11}}
\end{figure}

The growth rates of all these modes increase with the EP $\beta_f$, but the mode frequencies are different for different toroidal mode numbers (Fig.\ \ref{Fig11}). For $n=2$ RSAE, as the EP $\beta_f$ increases, the mode frequency mainly decreases and approaches to the lower boundary of the TAE gap. For $n=3$ EPM, the mode frequency decreases and touches the bottom line of the TAE gap more closely. In other words, stronger driving of energetic particle can overcome the continuum damping more easily. For $n=5$ case, the mode frequency increases first and then decreases in a narrow frequency region with EP $\beta_f$. For $n=6$ case, the mode frequency increases gradually with EP $\beta_f$.

\subsection{AE for $n=4$}
\subsubsection{EP $\beta$ fraction effect}
\begin{figure}
\includegraphics[height=4cm,width=6cm]{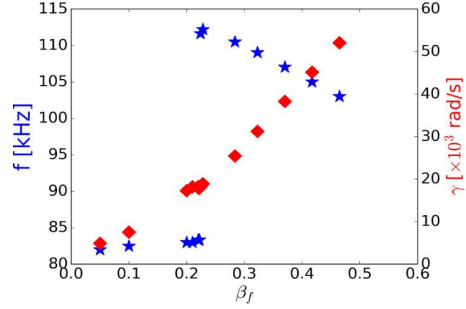}
\caption{The frequency (blue star)
and growth rate (red square) of TAE mode as functions of the $\beta$ fraction of energetic particles for $n=4$ mode. \label{Fig9}}
\end{figure}

For the $\beta$ fraction of energetic particle $\beta_f=0.2835$, the mode frequency is found to be $110.52kHz$, which is different from the frequency obtained from either GTAW or MEGA.
Then we reduce the $\beta$ fraction of energetic particle and find that the frequency drops to $83.33kHz$, which is close to the frequency from both GTAW and MEGA calculations. It can be seen from Fig.\ \ref{Fig9} that with the increase of EP beta fraction, the mode frequency jumps from a lower branch to a higher branch, whereas the corresponding growth rate increases monotonically.

\subsubsection{Mode structure}
\begin{figure}
\includegraphics[height=5cm,width=4cm]{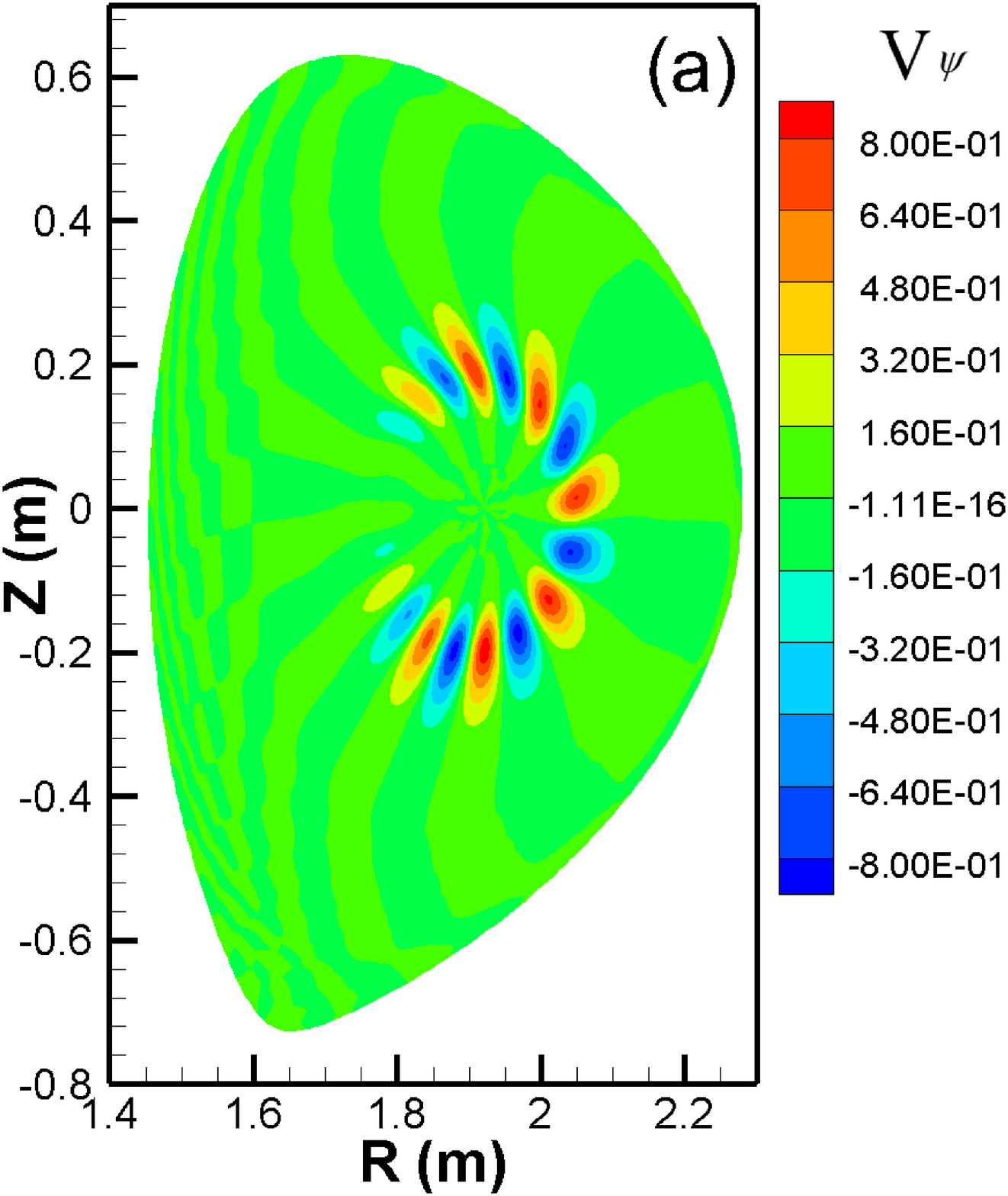}
\includegraphics[height=5cm,width=4cm]{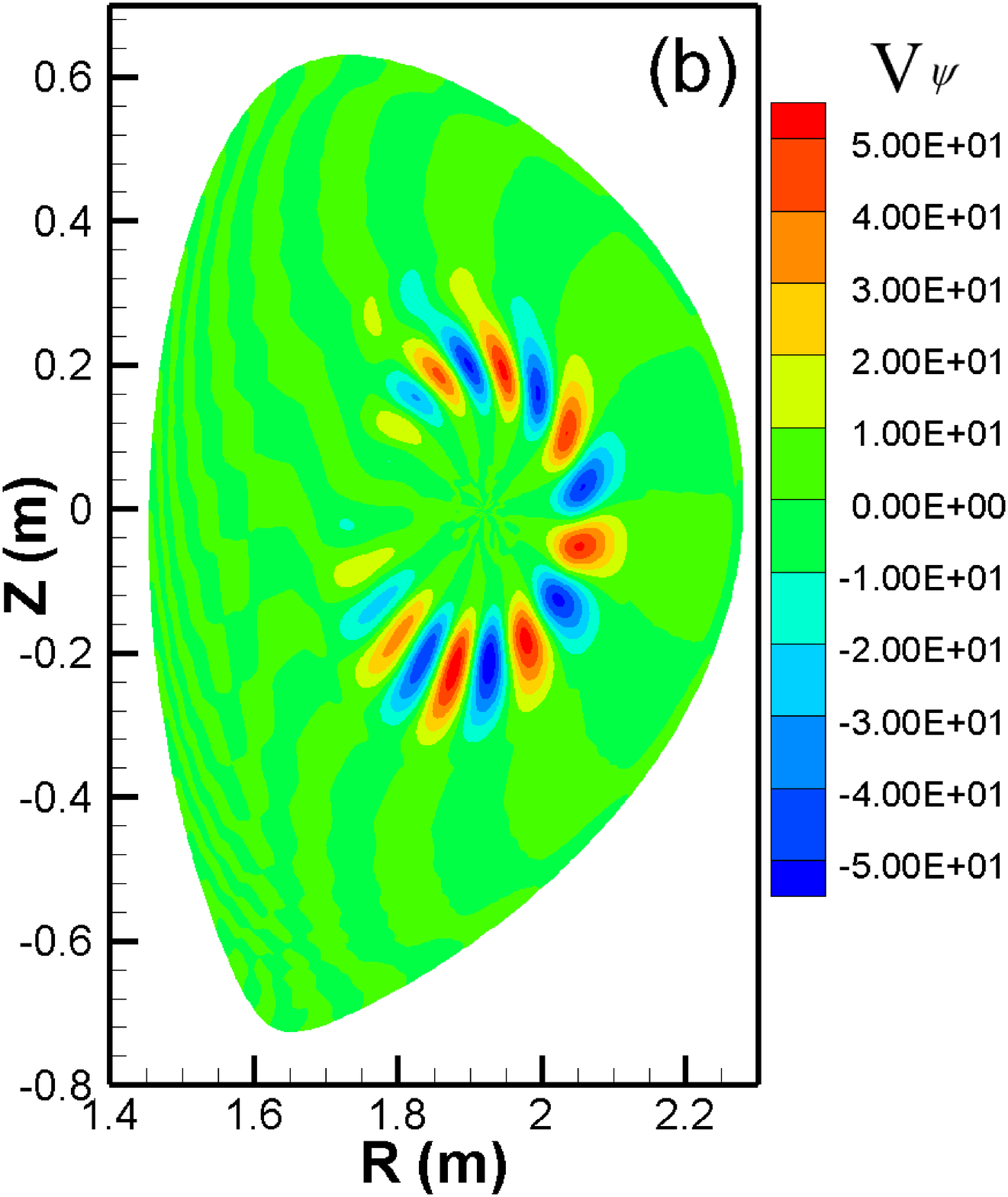}
\includegraphics[height=5cm,width=4cm]{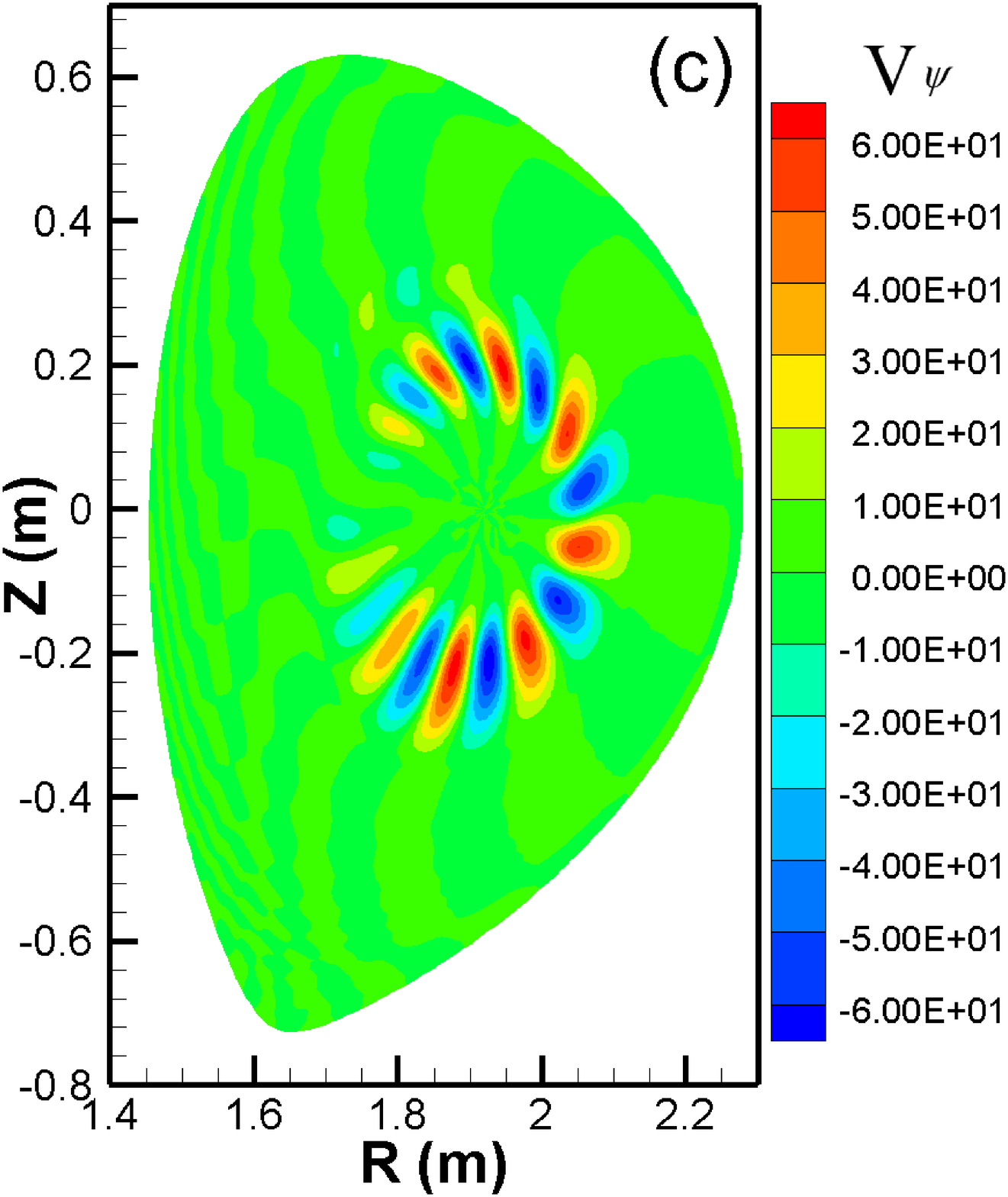}
\includegraphics[height=5cm,width=4cm]{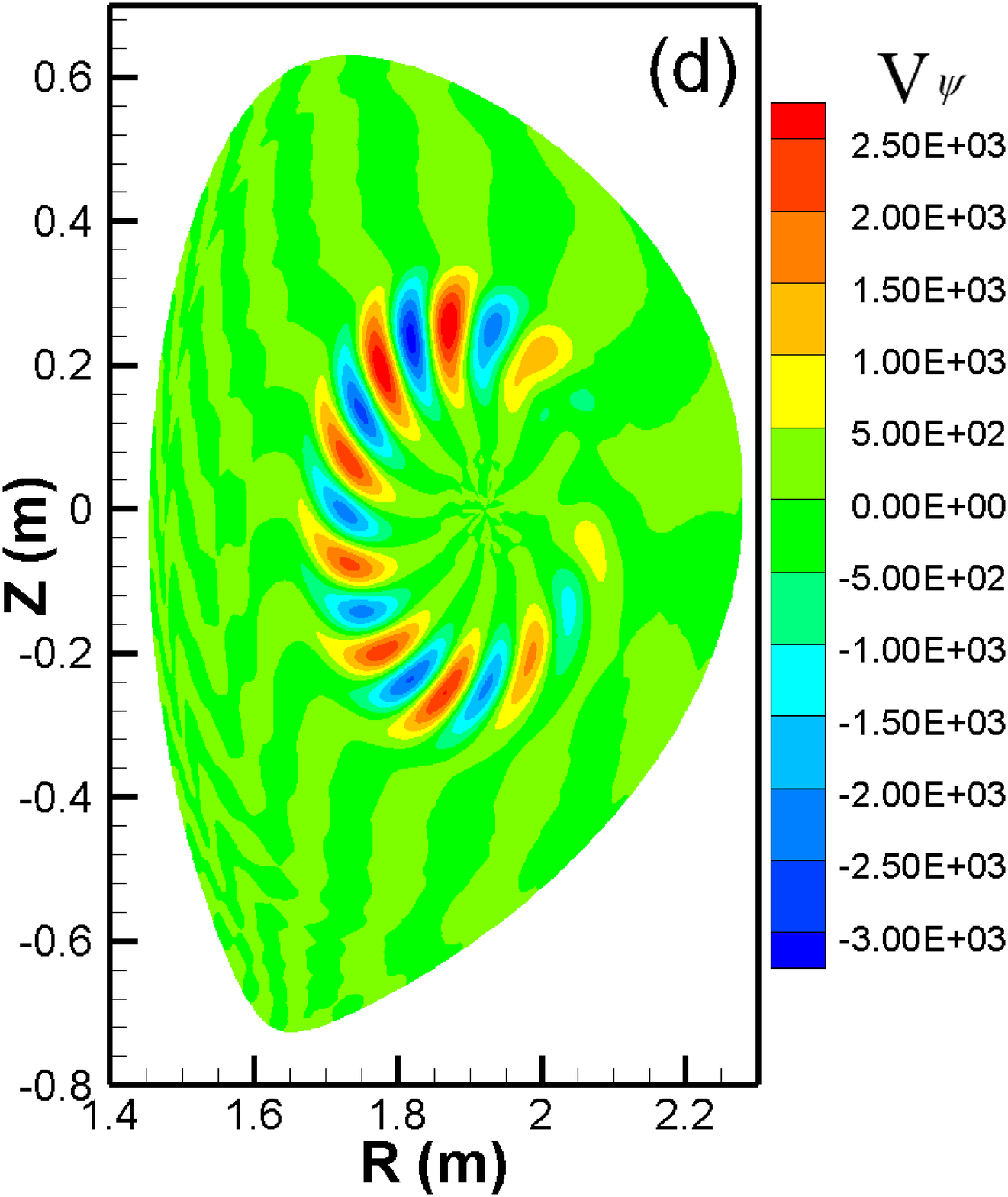}
\caption{Contour plot of the radial component of velocity of background plasma for toroidal mode number $n=4$ with different $\beta$ fraction of energetic particles. (a) $\beta_f=0.1$, (b) $\beta_f=0.222$, (c) $\beta_f=0.224$, (d) $\beta_f=0.4646$. \label{Fig8}}
\end{figure}

To identify the mode nature of these two branches, we examine the $2D$ mode structure with different EP $\beta$ fraction (Fig.\ \ref{Fig8}). When EP beta fraction is relatively small, the mode has a ballooning structure (Fig.\ \ref{Fig8} (a) and (b)). When EP $\beta$ fraction is relatively larger, the mode changes to an anti-ballooning structure (Fig.\ \ref{Fig8} (c) and (d)). All these modes are located in the radial region $0.2 < \sqrt{\psi} < 0.4$. These modes consist of poloidal harmonics $m=9$ and $m=10$.

\subsubsection{Mode identification}
\begin{figure}
\includegraphics[height=6cm,width=8cm]{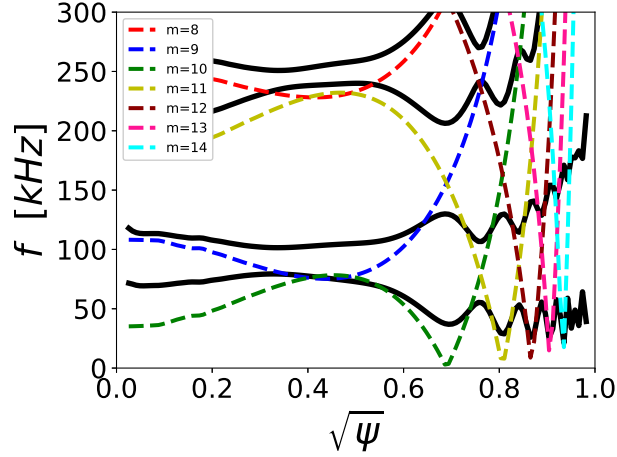}
\caption{Alfv\'en continua with toroidal mode number $n=4$ calculated based on the slow-sound approximation from AWEAC (solid lines). Alfv\'en continua from the cylindrical geometry limit (dotted lines) are also given. \label{Fig7}}
\end{figure}

From the $n=4$ Alfv\'en continua in Fig.\ \ref{Fig7}, we find that in the radial region $0.2 < \sqrt{\psi} < 0.4$, both lower and higher frequencies obtained from NIMROD simulations are within the TAE gap. Strong coupling between the $m=9$ and the $m=10$ harmonics induces the TAE gap, due to the fact that the two harmonics in the cylindrical limit intersect with each other. The ballooning and anti-ballooning mode structures of the two branches with different frequencies suggests that they may be identified as even and odd TAEs, which can be further confirmed in next two subsections.

\subsubsection{Mode transition between even and odd TAEs}
To investigate the mode transition between even and odd TAEs, especially the EP $\beta$ fraction threshold, we Fourier transform the time evolution of $B_R$, i.e., the radial component of magnetic field along the major radius. With EP $\beta$ fraction $\beta_f=0.05$, the mode frequency is $82kHz$ and the mode can be identified as even TAE. With EP $\beta$ fraction $\beta_f=0.4646$, the mode frequency is $103kHz$ and the mode can be identified as odd TAE.
For EP $\beta$ fraction $\beta_f=0.222$ and $\beta_f=0.224$, both even TAE and odd TAE coexist. That means there must be a threshold of EP beta fraction at which even TAE and odd TAE coexist with equal amplitudes. Usually, it's more difficult to excite the odd TAE, which can only be excited with larger $\beta$ fraction of energetic particles.

\begin{figure}
\includegraphics[height=4cm,width=4cm]{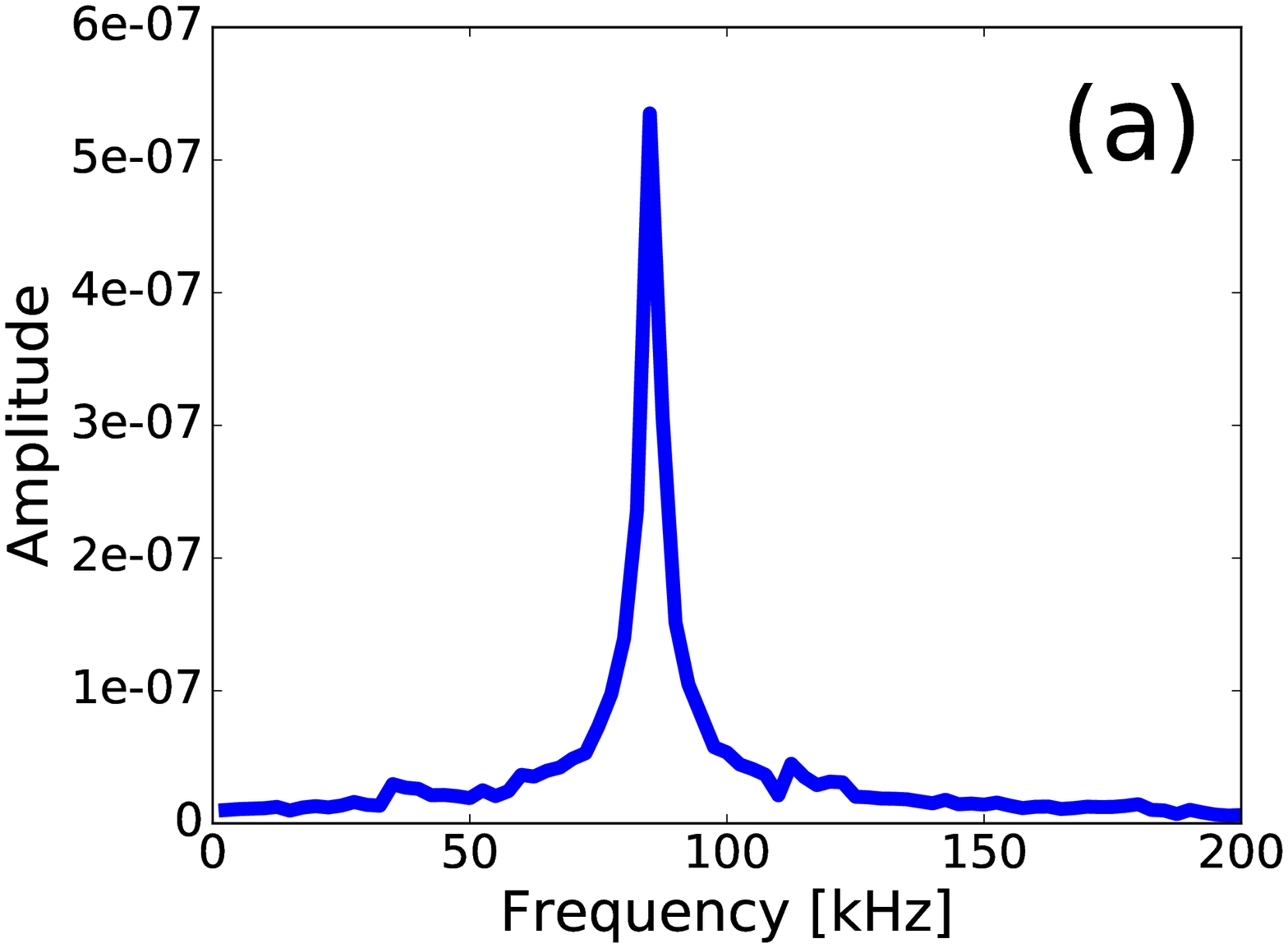}
\includegraphics[height=4cm,width=4cm]{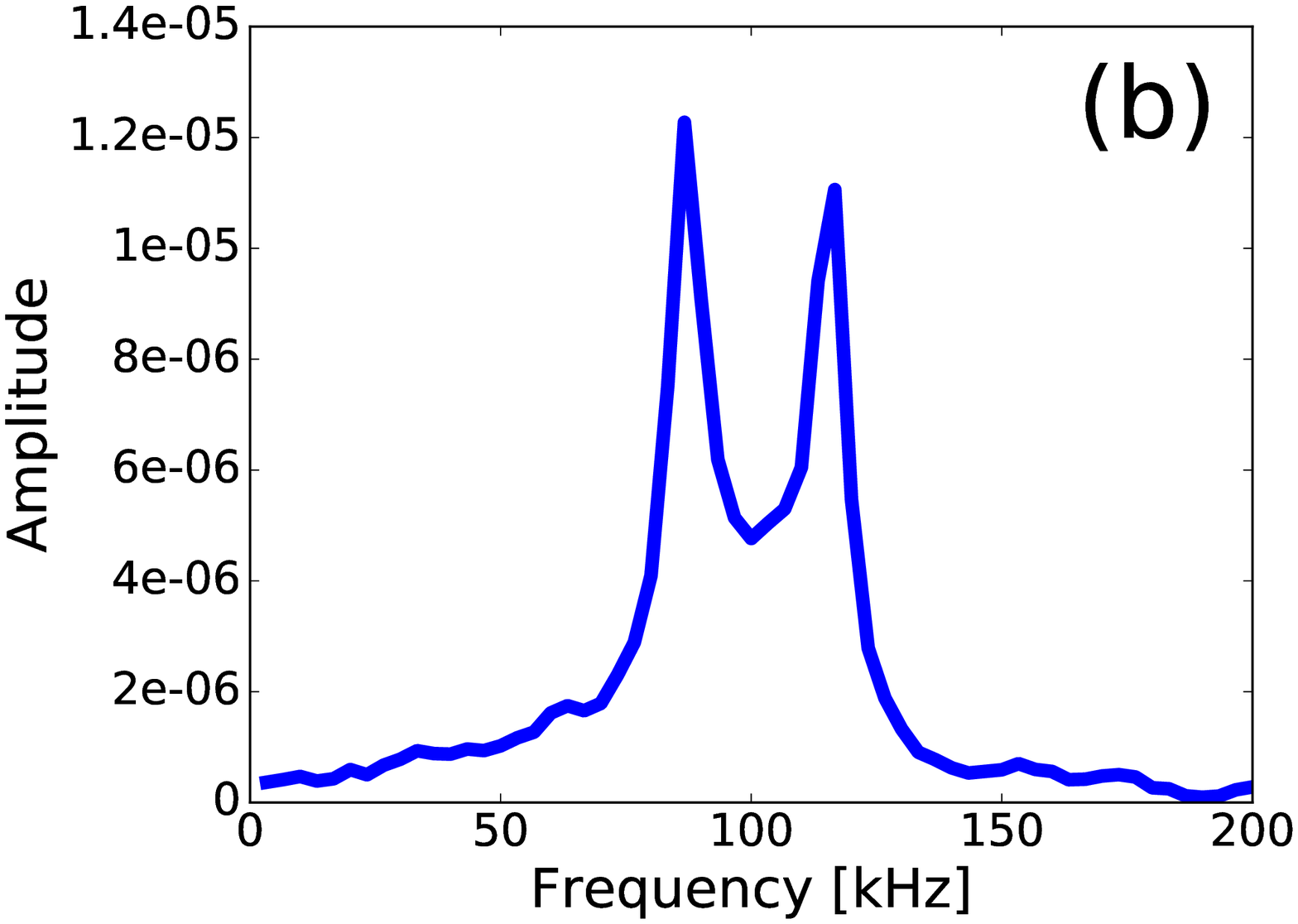}
\includegraphics[height=4cm,width=4cm]{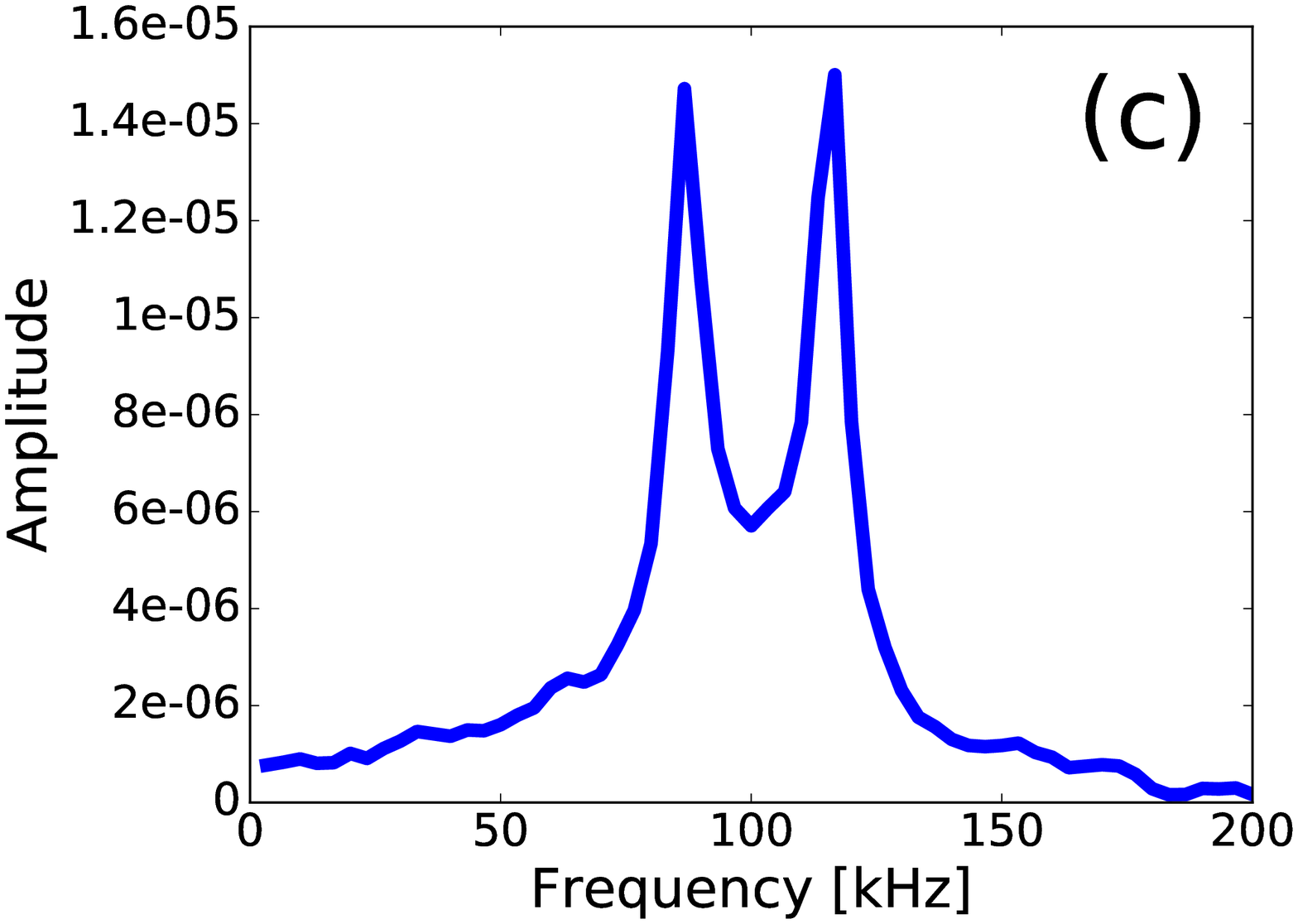}
\includegraphics[height=4cm,width=4cm]{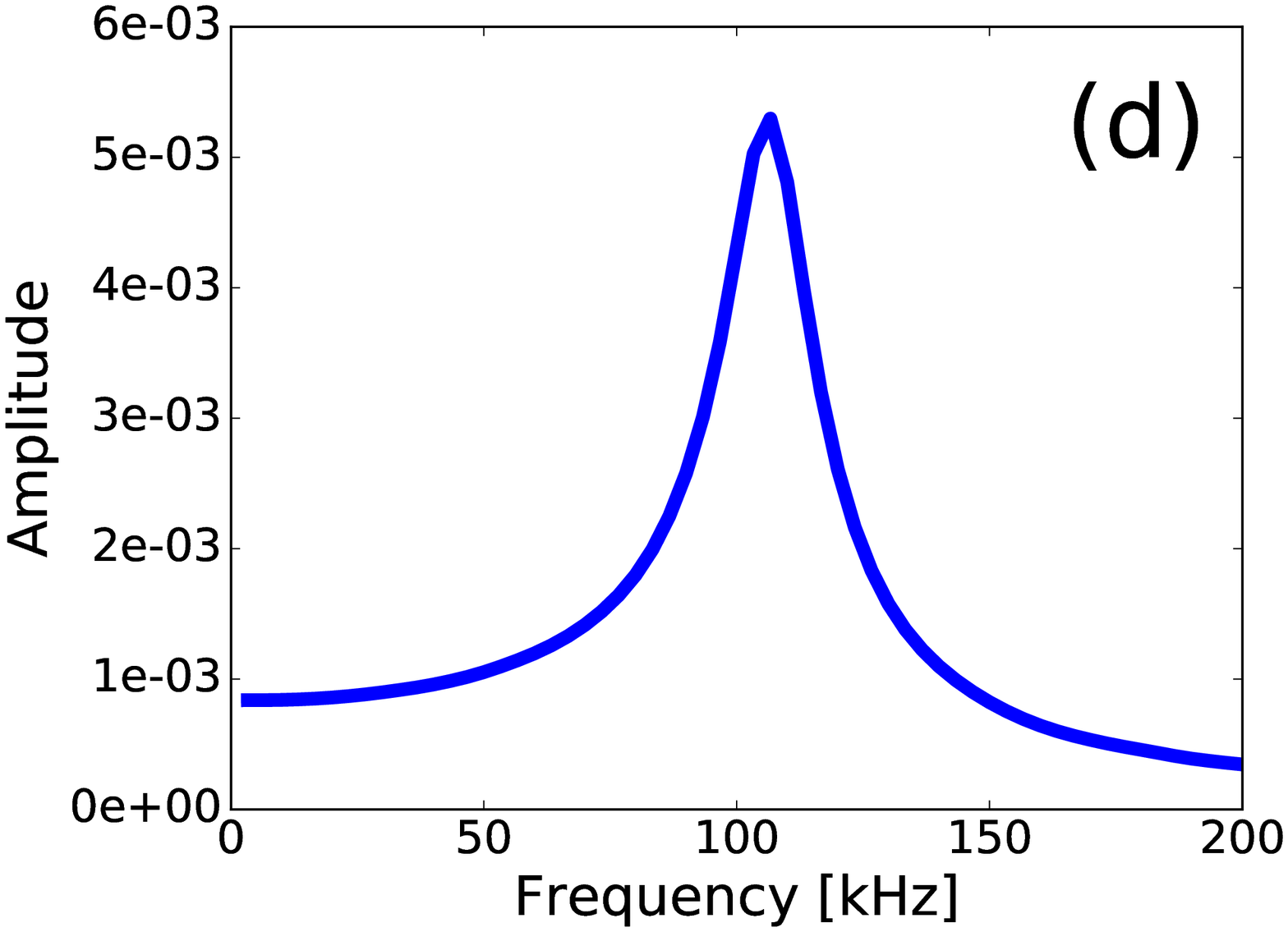}
\caption{The fourier spectrum of time evolution of $B_R$ for toroidal mode number $n=4$ with different $\beta$ fraction of energetic particles. (a) $\beta_f=0.05$, (b) $\beta_f=0.222$, (c) $\beta_f=0.224$, (d) $\beta_f=0.4646$.. \label{Fig10}}
\end{figure}

\subsubsection{Poloidal harmonic analysis}
\begin{figure}
\includegraphics[height=4cm,width=4.5cm]{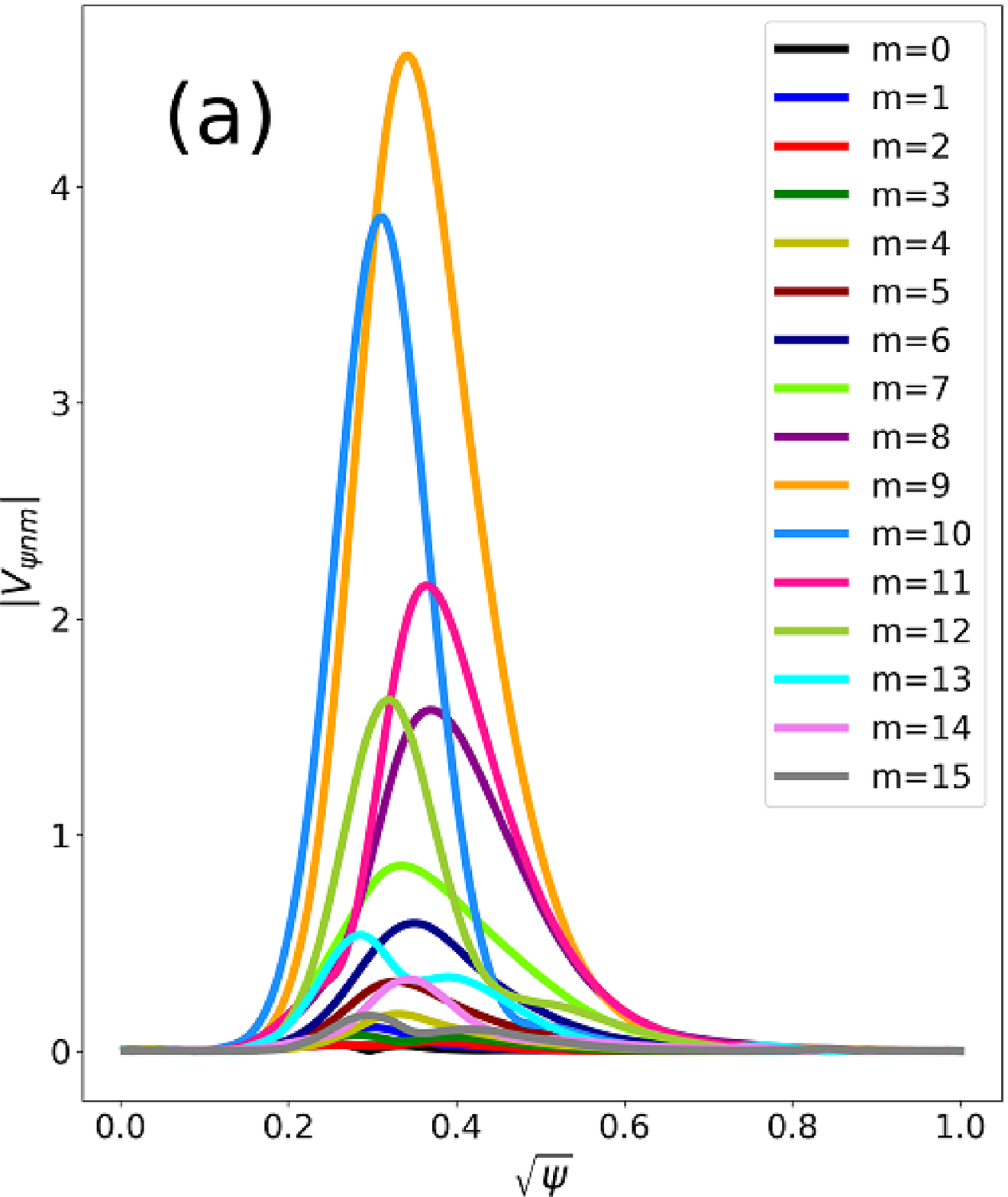}
\includegraphics[height=4cm,width=4.5cm]{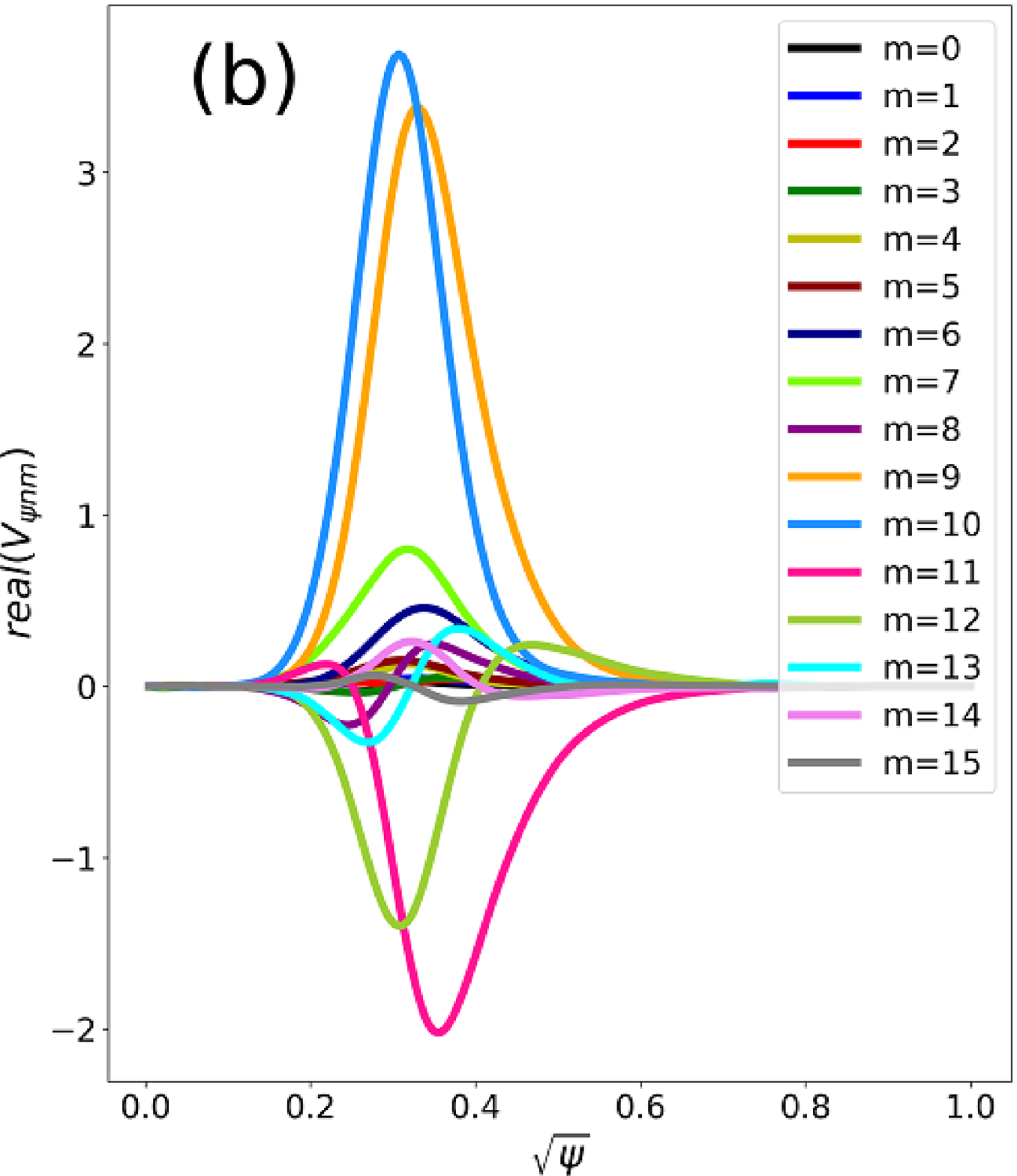}
\includegraphics[height=4cm,width=4.5cm]{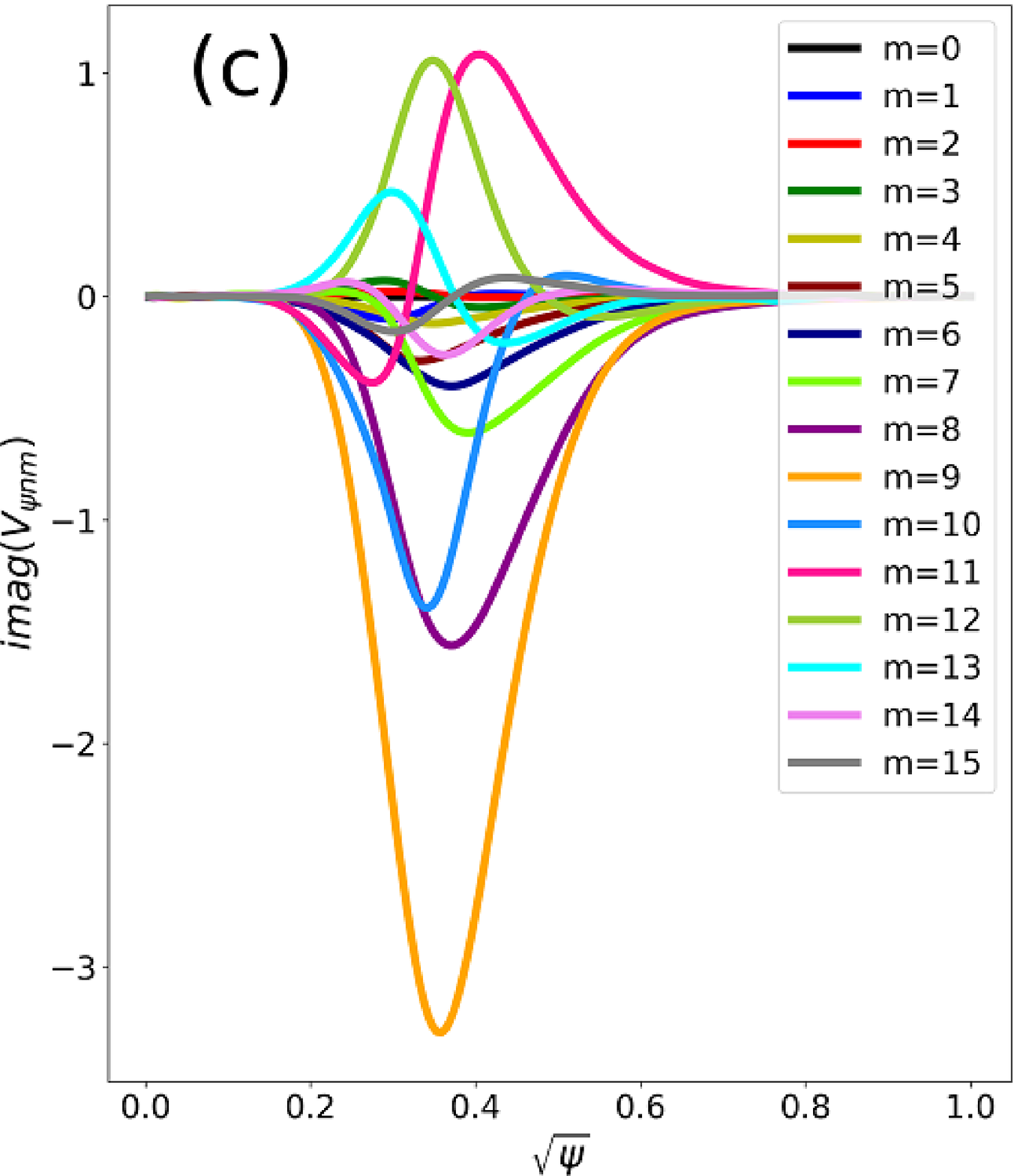}
\caption{The radial profiles of (a) amplitude of $|V_{\psi nm}|$, (b) real part of $V_{\psi nm}$, (c) imaginary part of $V_{\psi nm}$
for the $\beta_f=0.1$. \label{Fig12}}
\end{figure}

To distingusih odd and even TAEs, we further analyze the poloidal harmonics of the radial component of velocity.
The radial component of velocity is expanded in Fourier harmonics over $\phi$ as
\begin{equation}
V_\psi(\psi,\theta,\phi,t)= \sum_{n=-\infty}^{\infty} V_{\psi n}(\psi,\theta,t)e^{in\phi}
\simeq V_{\psi 0}(\psi,\theta,t)+ (\sum_{n=1}^{N} V_{\psi n}(\psi,\theta,t)e^{in\phi}+c.c.).
\end{equation}
In the analysis, we further take the Fourier transform of toroidal Fourier harmonics over $\theta$
\begin{equation}
V_{\psi n}(\psi,\theta,t)= \sum_{m=-\infty}^{\infty} V_{\psi nm}(\psi,t)e^{im\theta}
\simeq \sum_{m=-M}^{M-1} V_{\psi nm}(\psi,t)e^{im\theta} .
\end{equation}
When the energetic particle $\beta_f=0.1$ (Fig.\ \ref{Fig12} (a)), the poloidal harmonic $m=9$ is dominant and coupled with $m=10$ to generate TAE.
Both the $m=9$ and the $m=10$ harmonics have the same positive or negative sign in Fig.\ \ref{Fig12} (b) and (c), which is the feature of an even TAE.

\begin{figure}
\includegraphics[height=4cm,width=4.5cm]{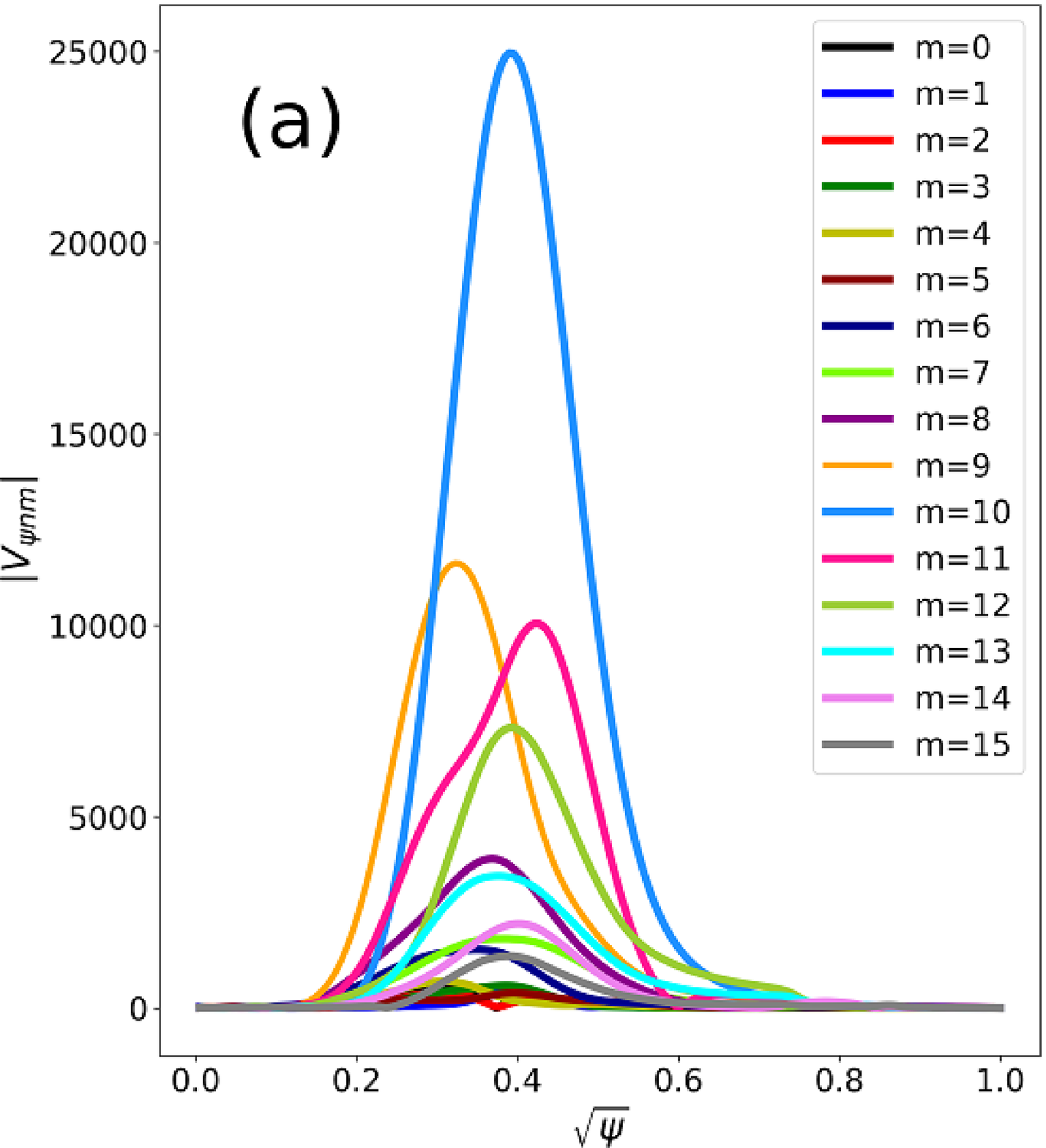}
\includegraphics[height=4cm,width=4.5cm]{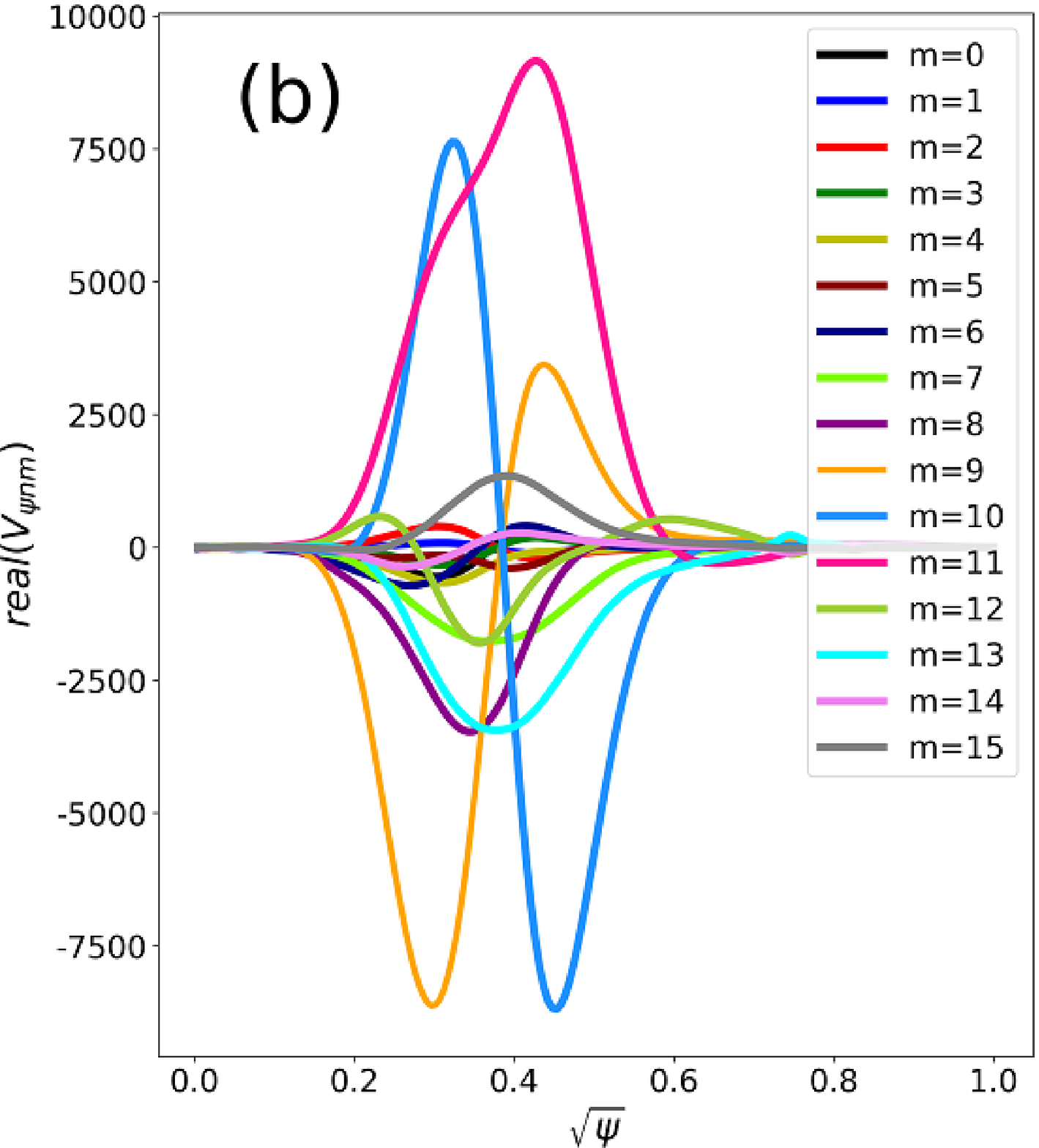}
\includegraphics[height=4cm,width=4.5cm]{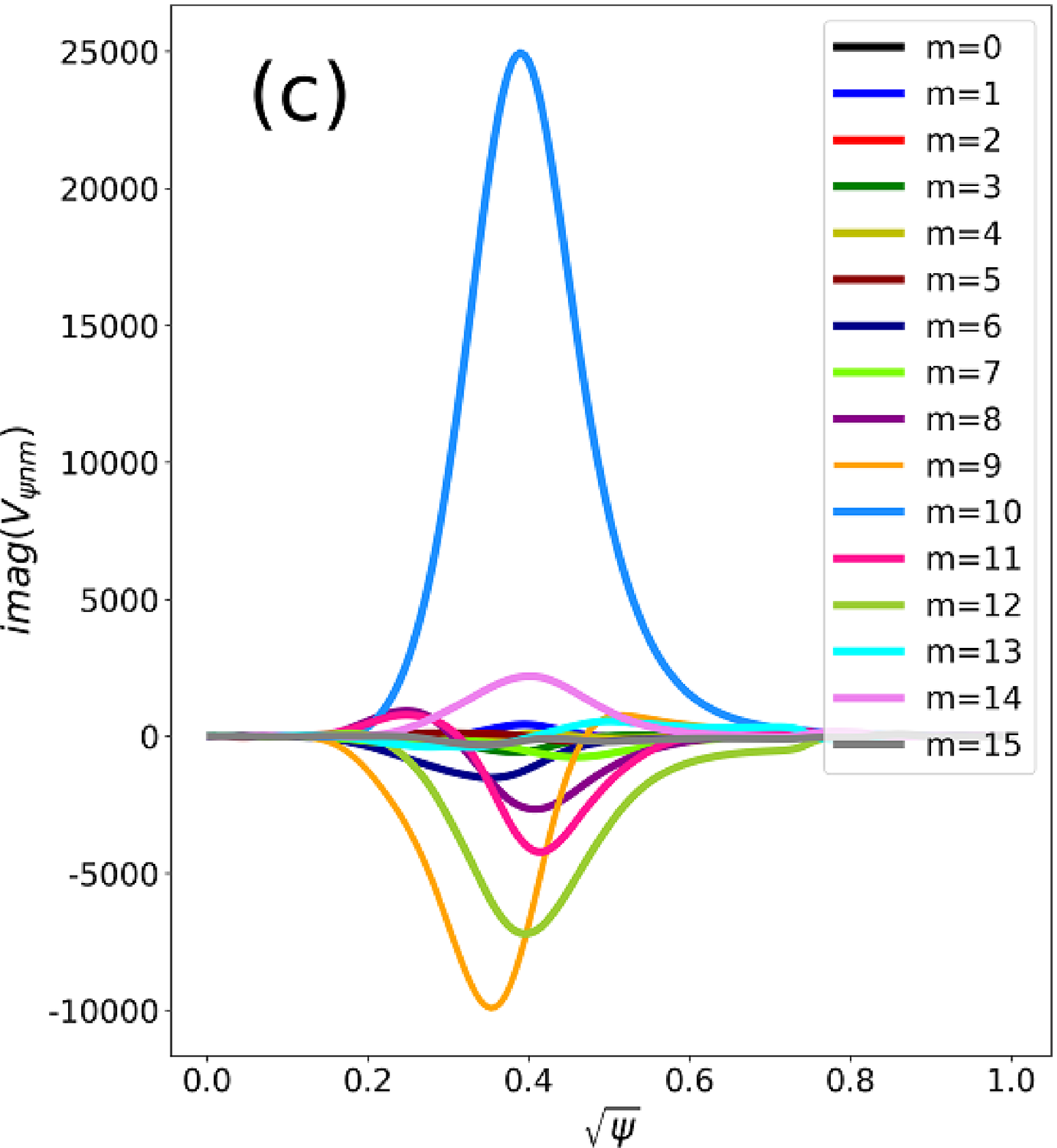}
\caption{The radial profiles of (a) amplitude of $|V_{\psi nm}|$, (b) real part of $V_{\psi nm}$, (c) imaginary part of $V_{\psi nm}$
for the $\beta_f=0.4646$. \label{Fig13}}
\end{figure}

When the energetic particle $\beta_f$ increases to $0.4646$ (Fig.\ \ref{Fig13} (a)), the poloidal harmonic $m=10$ becomes dominant and coupled with the $m=9$ harmonic to generate TAE. Now the $m=10$ and the $m=9$ harmonics have the opposite positive or negative sign in Fig.\ \ref{Fig13} (b) and (c), which is the feature of an odd TAE. As can be seen in Fig.\ \ref{Fig12} (a) and Fig.\ \ref{Fig13} (a), the peak value of the dominant mode is located at the radial position $\sqrt{\psi}=0.4$, where the $q_{min}$ is located. This is consistent with the even and odd TAE theory\cite{Fu1995a,Berk1995,Fu1995b}, according to which these TAEs are called core-localized TAEs.

\section{Summary and discussion}
Energetic particle driven modes on EAST are investigated using eigenvalue analysis and kinetic-MHD simulation. It is found that TAE/RSAE/EPM can be excited in the weak reverse shear equilibrium with various EP $\beta$ fraction for different toroidal mode numbers. The NIMROD simulation results have been successfully benchmarked with eigen-analysis (AWEAC and GTAW) and kinetic-MHD simulation (MEGA) with good agreements. Besides, a transition between even and odd TAEs due to enhanced driving of energetic particles is found in the NIMROD simulation, which is absent from previous simulation studies\cite{Hu2014,Hu2016}. As pointed out by Kramer et al\cite{Kramer2004}, a weak shear region in the core plasma and a flat central pressure profile are the two existence conditions for odd TAEs, which are satisfied in EAST discharge $\#48916$. Our results also suggest a new experimental scheme for identifying even and odd TAEs on EAST, where the frequencies of even and odd TAE would sweep downward and upward respectively with the decrease of EP driving (Fig.\ \ref{Fig9}). Although our work only focuses on the eigen-analysis and linear simulation, it provides a necessary foundation for nonlinear simulations including the wave-particle interactions, especially in phase space, in future.

\begin{acknowledgments}
This work was supported by the National Natural Science Foundation of China grant No. 11875253, the Fundamental Research Funds for the Central Universities under Grant Nos. WK3420000004, the National Magnetic Confinement Fusion Science Program of China under Grant Nos. 2014GB124002 and 2015GB101004, the National Key Research and Development Program of China No. 2017YFE0300500, 2017YFE0300501, the U.S. Department of Energy under Grant Nos. DE-FG02-86ER53218 and DE-SC0018001, and the 100 Talent Program of the Chinese Academy of Sciences. This research used the computing resources from the Supercomputing Center of University of Science and Technology of China, and the National Energy Research Scientific Computing Center, a DOE Office of Science User Facility supported by the Office of Science of the U.S. Department of Energy under Contract No. 13DE-AC02-05CH11231.
\end{acknowledgments}

\bibliography{EAST_TAE_BMref}


\end{document}